\PassOptionsToPackage{square,comma,numbers,sort&compress,super}{natbib}
\documentclass[twocolumn, aps,superscriptaddress,showpacs,floatfix]{revtex4-1}
\usepackage{multirow}
\usepackage{array}

\usepackage{achemso}

\usepackage{times}

\usepackage{color, soul}

\usepackage[table]{xcolor}
\usepackage{amsmath}
\usepackage{amssymb}
\usepackage{bm}
\usepackage{color}
\usepackage{graphicx}
\usepackage[latin1]{inputenc}
\usepackage[english]{babel}
\usepackage{color, soul}
\usepackage{ulem}
\usepackage{hyperref}
\hypersetup{
 colorlinks=true,
 linkcolor=blue,
 filecolor=magenta, 
 urlcolor=cyan,
}

\begin{document}
\newcommand{\smc}{\mathrm{g}_{\uparrow \downarrow}}

\newcommand{\alphaeff}{\alpha_{\rm eff}}

\newcommand{\Hk}{H_{\rm k}}

\newcommand{\Meff}{M_{\rm eff}}

\newcommand{\Ms}{M_{\rm \textrm{s}}}

\newcommand{\tFM}{t_{\rm FM}}

\newcommand{\tHM}{t_{\rm HM}}

\newcommand{\Ho}{\Delta H_{\rm 0}}

\newcommand{\Ks}{K_{\rm s}}

\newcommand{\Vdc}{V_{\rm dc}}

\newcommand{\Vsym}{V_{\rm sym}}

\newcommand{\Vasym}{V_{\rm asym}}

\newcommand{\DelH}{\Delta H}

\newcommand{\Hr}{H_{\rm \textrm{r}}}

\newcommand{\HR}{H_{\rm \textrm{R}}}

\newcommand{\jc}{j_{\rm \textrm{C}}}

\newcommand{\js}{j_{\rm s}}

\newcommand{\Lr}{\lambda_{\textrm{IREE}}}

\newcommand{\alphaR}{\alpha_{\textrm{R}}}

\newcommand{\TPSH}{\theta_\textrm{PSH}}

\newcommand{\IC}{I_{\rm C}}

\title{Large spin-to-charge conversion at the two-dimensional interface of transition metal dichalcogenides and permalloy}

\author{Himanshu Bangar}
\affiliation{
Department of Physics, Indian Institute of Technology Delhi, Hauz Khas, New Delhi-110016, India}%
\author{Akash Kumar}
\affiliation{
Department of Physics, Indian Institute of Technology Delhi, Hauz Khas, New Delhi-110016, India}%
\affiliation
{Department of Physics, University of Gothenburg, Gothenburg-412 96 , Sweden}
\author{Niru Chowdhury}
\affiliation
{Department of Physics, Indian Institute of Technology Delhi, Hauz Khas, New Delhi-110016, India}
\author{Richa Mudgal}
\affiliation
{Department of Physics, Indian Institute of Technology Delhi, Hauz Khas, New Delhi-110016, India}
\author{Pankhuri Gupta}
\affiliation
{Department of Physics, Indian Institute of Technology Delhi, Hauz Khas, New Delhi-110016, India}
\author{Ram Singh Yadav}
\affiliation
{Department of Physics, Indian Institute of Technology Delhi, Hauz Khas, New Delhi-110016, India}
\author{Samaresh Das}
\affiliation
{Center for Applied Research in Electronics, Indian Institute of Technology Delhi, Hauz Khas, New Delhi-110016, India}

\author{P. K. Muduli{$^{1,}$}}%
 \email{muduli@physics.iitd.ac.in}


\begin{abstract}
   Spin-to-charge conversion is an essential requirement for the implementation of spintronic devices.
  Recently, monolayers of semiconducting transition metal dichalcogenides (TMDs) have attracted considerable interest for spin-to-charge conversion due to their high spin-orbit coupling and lack of inversion symmetry in their crystal structure. However, reports of direct measurement of spin-to-charge conversion at TMD-based interfaces are very much limited. Here, we report on the  room temperature observation of a large spin-to-charge conversion arising from the interface of Ni$_{80}$Fe$_{20}$ (Py) and four distinct large area ($\sim 5\times2$~mm$^2$) monolayer (ML) TMDs namely,  MoS$_2$, MoSe$_2$, WS$_2$, and WSe$_2$. We show that both spin mixing conductance and the Rashba efficiency parameter ($\Lr$) scales with the spin-orbit coupling strength of the ML TMD layers. The $\Lr$ parameter is found to range between $-0.54$ and $-0.76$~nm for the four monolayer TMDs, demonstrating a large spin-to-charge conversion. Our findings reveal that TMD/ferromagnet interface can be used for efficient generation and detection of spin current, opening new opportunities for novel spintronic devices. 
\end{abstract}

\maketitle
\section{INTRODUCTION}
Spintronics offers a route to fulfill the increasing demand for faster processing hardware and mobile applications in the age of the internet of things.~\cite{liu2019overview,jung2022crossbar} One of the key requirements in spintronics is the efficient generation and detection of spin currents, which was traditionally accomplished by the use of ferromagnetic (FM) layers.
Recent research has focused on utilizing spin-orbit interaction of heavy metal (HM) and FM heterostructures for spin-to-charge conversion.~\cite{gambardella2011current,liu2012spin,Manchon2019review,kumar2021large} This has led to a growing field of spin-orbitronics in which coupled interaction of spin and orbital momentum of the electron is employed for interplay between charge and spin currents~\cite{sinova2015spin,Manchon2019review}. When an electric current is passed through the HMs with strong spin-orbit coupling (SOC), it produces a transverse spin current due to the spin Hall effect (SHE)~\cite{dyakonov1971current,hirsch1999spin,sinova2015spin,Manchon2019review}. Conversely, when a spin current is applied to a HM, charge current can be generated via inverse SHE (ISHE). Spin currents have been used in a number of spintronic applications, such as  domain wall motion~\cite{khvalkovskiy2013matching}, spin-torque switching~\cite{liu2012spin} and spin Hall nano-oscillators~\cite{chen2016spin}. 
For optimum performance of these spin-orbitronics based devices, high spin-to-charge conversion efficiency is essential  \cite{liu2012spin,kumar2021large}. Recent years have seen a surge of interest in two-dimensional (2D) interfaces for spin-to-charge interconversion via the  Rashba-Edelstein effect (REE)~\cite{bychkov1984properties,rashba2015symmetry,edelstein1990spin,ando2017spin,shen2014microscopic, sanchez2013spin,manchon2015new,rojas2016spin,kurebayashi2022magnetism}. The REE mechanism has the potential to be more efficient than the SHE mechanism in three-dimensional (3D) HMs.\cite{ando2017spin,vaz2019mapping}

Spin-charge interconversion via REE occurs at interfaces with broken spatial inversion symmetry due to the spin-orbit interaction and built-in electric potential in the direction normal to the film surface~\cite{gambardella2011current,rezende2020fundamentals,Manchon2019review}. Fig.~\ref{fig:1}A,~\cite{rashba2015symmetry} shows the Rashba-type spin splitting of the energy dispersion that arises at the interface. It is characterized by (i) spin splitting of bands and (ii) spin-momentum locking, which leads to two Fermi contours with opposite helicity, as shown by the dashed lines in Fig.~1B. The Rashba Hamiltonian, $\HR$ is expressed as:
$\HR=\alphaR\left(\bm{k}\times \bm{\hat{z}}\right).\hat{\bm{\sigma}}$,
where, $\bm{k}$ denotes the momentum vector, $\hat{\bm{\sigma}}$ denotes the Pauli spin matrix vector, $\bm{\hat{z}}$ denotes the unit vector normal to the interface and $\alphaR$ is the Rashba coefficient~\cite{bychkov1984properties,ando2017spin}. The energy dispersion depicted in Fig.~\ref{fig:1}A is a solution of the Rashba Hamiltonian.
The relation for the in-plane momentum, $k_{x}$ of the Rashba interface is given as \cite{ando2017spin}:
\begin{figure*}[t!]
\centering
\includegraphics[width=17cm]{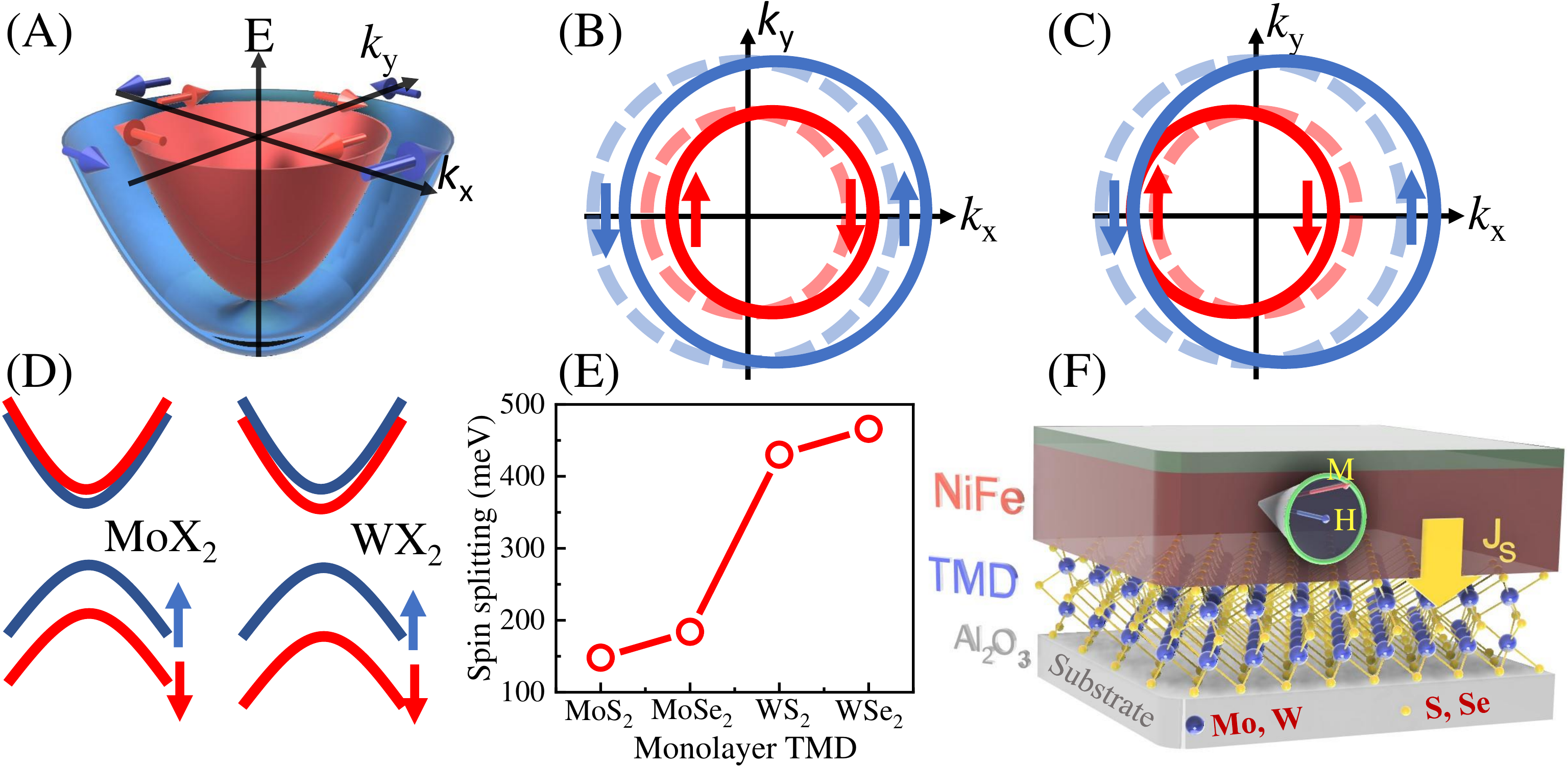}
\caption{\label{fig:1} \textbf{Schematic representation of the REE, IREE effects, and key features of ML TMDs.} \textbf{(A)} A typical energy dispersion curve for the Rashba interaction. The inner and outer circles have an opposite spin texture at the Fermi level \cite{ando2017spin}. \textbf{(B)} Illustration of the shift in Fermi contours under the influence of a charge current along $-x-$direction leading to spin accumulation at the interface (REE effect). \textbf{(C)} Schematic representing the shift in Fermi contours due to injection of spin current with polarization along $+y-$direction leading to a charge current at the interface (IREE effect). \textbf{(D)} Schematic illustration of spin splitting in the valence band around the $K$-point of MoX$_2$ and WX$_2$ ML TMDs. Here, X denotes S or Se. \textbf{(E)} Theoretical spin splitting of the uppermost valence band at the $K$-point for different ML TMDs~\cite{liu2013three}. \textbf{(F)} The ML TMD/ferromagnet (FM) structure used in this work is depicted schematically, illustrating how spin current is injected via spin pumping.} 
\end{figure*}

\begin{equation}\label{Hamiltonian}
\begin{split}
E(k_x)=\frac{\hbar^2 k^2_x}{2m^*}\pm \alphaR k_x,
\end{split}
\end{equation}
where, $m^*$ denotes the effective mass of the electron. Hence, the magnitude of the spin splitting of bands caused by the Rashba interaction depends on $\alphaR$, which is dependent on the SOC at the interface. The Fermi contour of Fig.~\ref{fig:1}B shows that for $k_{x}>0$, the number of spin-up ($\uparrow$) carriers is larger than that of the spin-down ($\downarrow$) carriers. When an in-plane electric field is applied along the $-x-$direction, two Fermi contours (solid lines), which have opposite spin helicity, are shifted along the
$+x-$direction, resulting in a net spin
accumulation with spin polarization along the $+y-$direction. This is referred to as the
REE \cite{ando2017spin}. On the other hand, if the spin is injected into the Rashba interface with spin current having a polarization along $+y-$direction, then the two Fermi contours are shifted in opposite directions (Fig.~1C, solid lines), leading to a net charge current at the interface along the $-x-$direction. This phenomenon is referred to as the inverse REE (IREE), which is the Onsager reciprocal of the REE~\cite{shen2014microscopic}.

Spin-to-charge conversion utilizing IREE has received considerable interest after Rojas S\'{a}nchez \textit{et al.} reported a high spin-to-charge conversion efficiency at the silver (Ag)/bismuth (Bi) interface~\cite{sanchez2013spin}. Subsequently, several works have reported spin-to-charge conversion at the interface between two non-magnetic materials.~\cite{sangiao2015control,zhang2015spin,isasa2016origin,nomura2015temperature,karube2016experimental} IREE has also been used for the generation of THz radiation~\cite{zhou2018broadband}. The spin-to-charge conversion becomes very complex in a 3D FM-HM  system with large  REE  contributions due to entangled surface and bulk contributions~\cite{Manchon2019review}. Additionally, the error in estimating interfacial thickness leads to  important parameters being underestimated or overestimated. It is thus desirable to have a system that is truly 2D,  from which the extracted parameters can be accurately attributed to the IREE.  
However, the reported studies of spin-to-charge conversion in large area true 2D-based systems are very much limited.~\cite{ohshima2014observation, mendes2015spin}.

Recently, monolayers (ML) of semiconducting 2D materials such as the transition metal dichalcogenides (TMDs) has gained huge interest for spin-to-charge conversion owing to their (i) large SOC derived from the $d$-orbitals of the transition metal~\cite{gusakova2017electronic} and (ii) lack of inversion symmetry in their crystal structure~\cite{mendes2018efficient, bhowal2020intrinsic}. Both of these characteristics are necessary components of a Rashba interface. Recently, strong REE induced spin-orbit torques~\cite{shao2016strong}, unconventional charge-spin conversion and spin-orbit torques~\cite{zhao2020unconventional,macneill2017control} are reported in TMD/FM systems. It is also demonstrated that the TMDs enhances the spin-to-charge conversion efficiency significantly when used as an insertion layer between HM and FM layers.~\cite{lee2020enhanced,lee2020enhanced2,dastgeer2019distinct} Out of various studied TMDs, Mo and W-based TMDs stand out due to the presence of heavy transition metals. Moreover, the valence band of ML MoX$_2$ and WX$_2$ TMDs (X= S or Se) exhibits a spin splitting around the $K$-point due to the presence of a large SOC as schematically shown in Fig.~1D~\cite{liu2013three}. The theoretical spin splitting values of the valence band around the $K$-point for these ML TMDs are shown in Fig.~1E~\cite{liu2013three}, which allows the control of spin-to-charge conversion via SOC of the ML TMD~\cite{manchon2015new}.

Here, we report a systematic study of spin-to-charge conversion at the interface of Ni$_{80}$Fe$_{20}$ (Permalloy, abbreviated as Py hereafter) and large area ML TMDs namely MoS$_2$, MoSe$_2$, WS$_2$, and WSe$_2$ (Fig. 1F). To date, spin-charge interconversion studies on such systems are mostly performed in non-local geometry as growing TMDs with large area coverage is challenging~\cite{PhysRevResearch.2.033204,PhysRevResearch.2.013286}. We use large area chemical vapor deposited ML TMDs interfaced with sputtered Py thin films. TMD layers having different SOC allows us to control $\alphaR$ and hence spin-to-charge conversion. We observe a large spin-to-charge conversion in these systems having efficiencies that scale with SOC of the ML TMD layer consistent with the IREE effect. The magnitude of spin-to-charge conversion efficiency exceeds that of the 3D FM/HM systems with the maximum efficiency in W-based TMDs as compared to Mo-based TMDs. 

\section{RESULTS AND DISCUSSION}

\begin{figure*} [t!]
\centering
\includegraphics*[width=17cm]{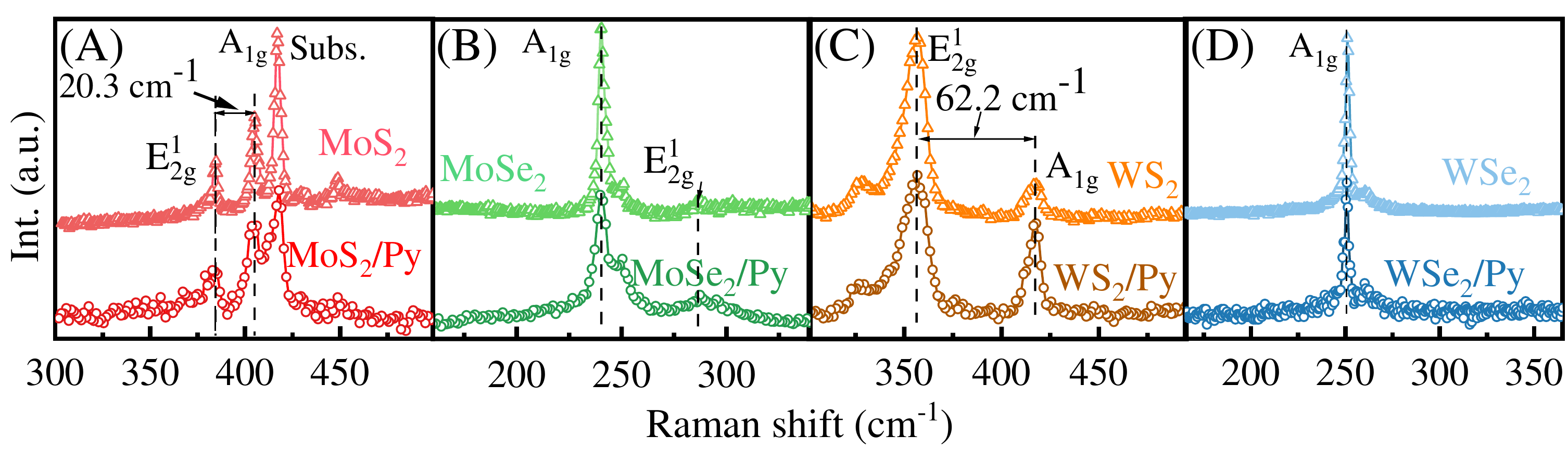}
\caption{\label{fig:2} \textbf{Raman measurements.} \textbf{(A-D)} Raman spectra for different ML TMDs before (open triangles) and after the deposition of Py and Ta layers (open circles) \textbf{(A)} MoS$_2$, \textbf{(B)} MoSe$_2$, \textbf{(C)} WS$_2$, and \textbf{(D)} WSe$_2$, respectively. }
\end{figure*}
We studied Al$_2$O$_3$/ML TMD/Py(10 nm)/Ta(3 nm) samples for four distinct large area ($\sim 5\times2$~mm$^2$) TMDs namely,  MoS$_2$, MoSe$_2$, WS$_2$, and WSe$_2$. A reference sample without the TMD layer was also prepared \textit{simultaneously} for comparison. Figure~2A-D shows the measured Raman spectra observed on pristine ML TMDs (open triangles) and after deposition of Py/Ta bilayer (open circles), respectively. Two peaks corresponding to E$_{\rm2g}^{1}$ and A$_{\rm1g}$ Raman modes were observed for MoS$_2$ (Fig.~2A) and WS$_2$ (Fig.~2C) samples.~\cite{kumar2020direct,bansal2019extrinsic} The separation of the E$_{\rm2g}^{1}$ and A$_{\rm1g}$ peaks is found to be 20.3 cm$^{-1}$ for MoS$_2$ and 62.2 cm$^{-1}$ for WS$_2$, confirming their ML thickness~\cite{shi2018assisted,berkdemir2013identification}. The absence of B$_{\rm2g}^{1}$ mode around 353 cm$^{-1}$ (Fig.~2B) and 309 cm$^{-1}$ (Fig.~2D) is consistent with the previously reported Raman spectra for ML MoSe$_2$ and WSe$_2$, respectively~\cite{tonndorf2013photoluminescence}. A single strong A$_{\rm1g}$ Raman mode at 240 cm$^{-1}$ (Fig. 2B) and 251 cm$^{-1}$ (Fig.~2D) are the fingerprints of ML MoSe$_2$ and WSe$_2$, respectively.~\cite{mahjouri2016tailoring} 
In order to confirm that we have large area coverage of ML TMDs, we performed Raman measurements at multiple locations and obtained very consistent Raman spectra (Fig.~S1 and S2), indicating that our TMD samples have ML thickness over the entire area of the sample ($5\times2$ mm$^2$). The peak positions of the Raman spectra obtained after the deposition of Py and Ta layers in Fig.~2 coincides with their corresponding pristine ML TMDs for all four cases. This demonstrates that the quality of the ML TMD  is not affected after the deposition of the Py and Ta layers. 
In addition, this reveals no diffusion of Mo, W, S, or Se into the Py layer, which would have adversely altered the Raman signal from the ML TMDs.
We also did not observe any degradation of the surface morphology of the Py/Ta stacks grown on top of the TMD, as shown in Supplementary Fig.~S3. This is because  of the (1) low atomic roughness ($<$0.15~nm) of all the TMD layers, and (2) the low working pressure for the growth of the Py layer. The Py deposition was optimized following our earlier work on the growth of Py on Graphene, where we obtained minimum defects in Graphene for room temperature deposition and low working pressure.~\cite{chaurasiya2019direct}

\begin{figure*} [t!]
\includegraphics*[width=15.5cm]{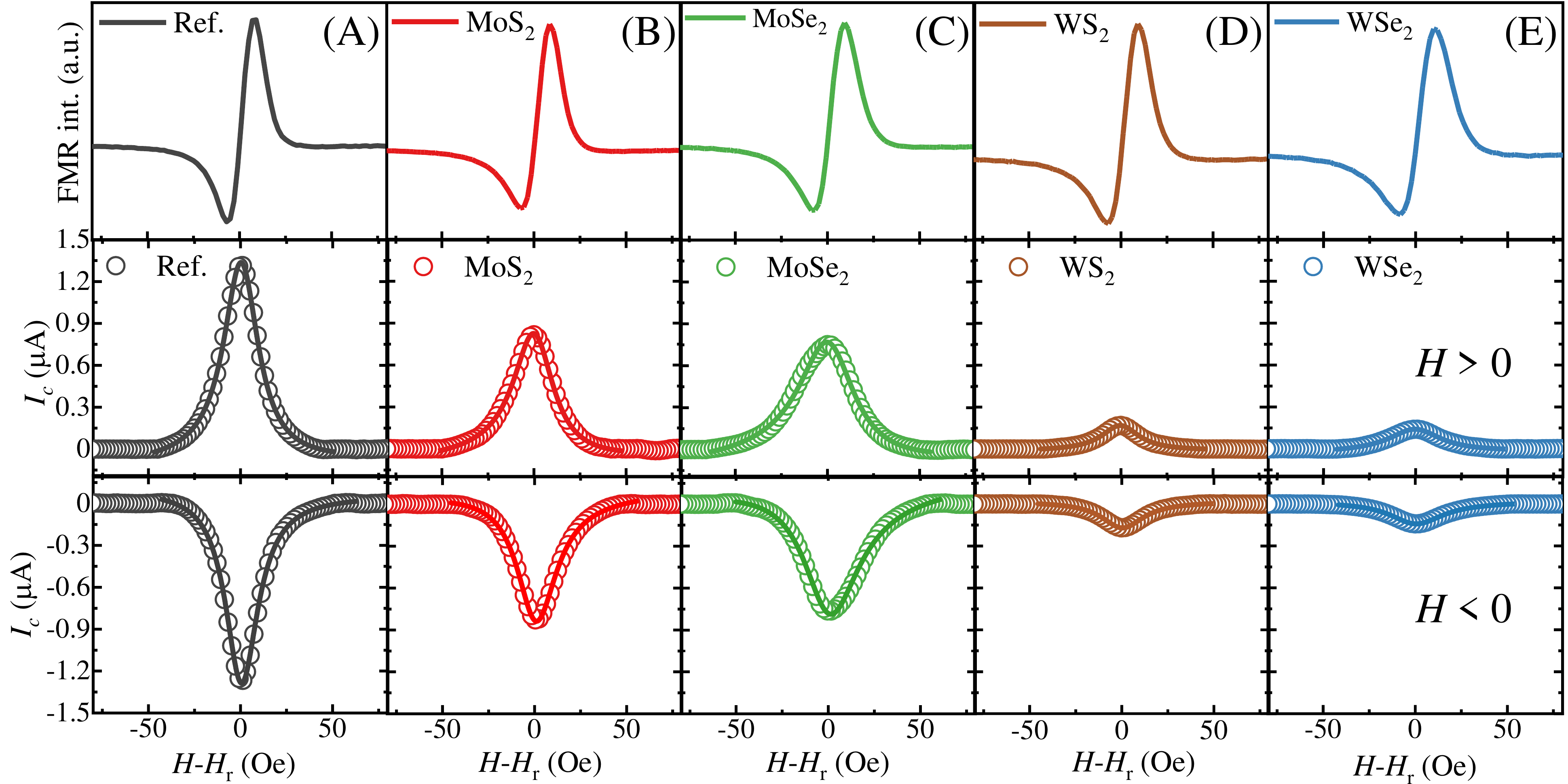}
\caption{\label{fig:3} \textbf{FMR and IREE measurements.} Field dependence of FMR (top row) and corresponding $\IC$ derived from IREE measurements (middle and bottom row) for \textbf{(A)} Py(10)/Ta(3), \textbf{(B)} ML MoS$_2$/Py(10)/Ta(3), \textbf{(C)} ML MoSe$_2$/Py(10)/Ta(3), \textbf{(D)} ML WS$_2$/Py(10)/Ta(3), \textbf{(E)} ML WSe$_2$/Py(10)/Ta(3) samples. Here, numbers in parenthesis reflect thickness in nanometers and ML is abbreviation for monolayer. The middle row is for for $H>0$ while the bottom row is for $H<0$. The solid lines in middle and bottom row are fits to experimental data.}
\end{figure*}

In order to measure the spin-to-charge conversion in our samples, we drive the Py layer to the ferromagnetic resonance (FMR). The upper panel of Fig.~3 shows the measured derivative of FMR absorption around the resonance field ($\Hr$) for various ML TMD/Py samples along with the reference sample (without TMD layer). The FMR of the magnetization of the Py layer leads to the pumping of the spin current to the ML TMD/Py interface, as shown schematically in Fig.~1F. The pumped spin current is subsequently converted to charge current and is measured as a voltage signal ($\Vdc$) using field modulation technique~\cite{kumar2018large}. The field dependence of charge current ($\IC$) around $\Hr$ corresponding to the measured $\Vdc$ is plotted in the middle panel ($H > 0$) and lower panel ($H < 0$) of Fig.~3. 
 
The measured  field dependence of $\IC$ is dominated by a symmetrical lineshape for all samples, including the reference sample. We fitted the measured spectrum using a combination of the symmetric and antisymmetric components of the Lorentzian function and found the antisymmetric component to be negligible, indicating the lack of signal due to the galvanic effects such as anomalous Hall effect (AHE) and/or anisotropic magnetoresistance (AMR)~\cite{lustikova2015vector, kumar2018large,cerqueira2018evidence}. In fact, our measurement geometry is optimized to minimize the galvanic effects by using a thin mica sheet (100~$\mu$m) between the sample and CPW, which minimizes RF electric field. 
Furthermore, we observe the expected sign reversal of $\IC$ when the field direction is reversed as shown in the middle and bottom row of Fig.~3. These observations indicated that the signal we observe is primarily due to spin-to-charge conversion.

The magnitude of $|\IC|$ is found to drop with respect to the reference sample as TMD layer is varied in the sequence MoS$_2\rightarrow$MoSe$_2\rightarrow$WS$_2\rightarrow$WSe$_2$ (second and third row of Fig.~3). Furthermore, the drop is stronger for W-based TMDs compared to the Mo-based TMDs. Since our system comprises of Al$_2$O$_3$/ML TMD/Py/Ta, $\IC$ can arise due to spin-to-charge conversion from (i) the ML TMD/Py interface, (ii) from the bulk of Py layer due to self-induced ISHE~\cite{gladii2019self} and (iii) from the ISHE of Ta capping layer as well other interfacial mechanisms from Py and Ta interface. We found all these contributions to be present in our measurements. The signal from the reference sample is due to the self-induced ISHE as explained in the supplementary material Sec.~S5. Hence, in order to extract the signal coming from ML TMD/Py interface, we subtracted the signal coming from the reference layer: $\IC^{TMD/Py}=\IC^{TMD/Py/Ta}-\IC^{Py/Ta}$. The 2D or the surface charge current density ($\jc$) corresponding to $\IC$ is given as $\jc=\frac{\IC}{b}$~\cite{zhang2015spin}. Here, $b$ is the width of the sample, which was the same for all the samples. Thus, we can determine the  2D charge current density of the ML TMD/Py interface $\jc^{TMD/Py}$, which is found to have a negative sign. The magnitude of $|\jc^{TMD/Py}|$ is found to increase monotonically (Fig.~4C) in the sequence MoS$_2\rightarrow$MoSe$_2\rightarrow$WS$_2\rightarrow$WSe$_2$, suggesting a direct correlation of the measured  $\jc^{TMD/Py}$ and SOC strength of the ML TMD. 

To further confirm that the signal originates at the ML TMD/Py interface, we also measure $\IC$ by inserting a Cu spacer layer between the ML TMD and the Py layer. We no longer observe the decrease in $\IC$  as shown in Fig.~S4 of supplementary material Sec.~S3. Cu was chosen as the insertion layer due to its extremely long spin diffusion length [$>$350 nm]~\cite{villamor2013contribution,PhysRevB.98.024416} and hence spin current should travel through it without dissipating. Thus, the decrease in $\IC$ in the case of ML TMD/Py samples in comparison to their reference sample in Fig.~3 can be attributed to the IREE at the ML TMD/Py interface rather than ISHE in the Py layer or Ta capping layer. As another confirmation of this behavior, and to understand the negative sign of $\IC^{TMD/Py}$, we also prepared a series of Py samples with various capping layers of Ta, Pt, and SiO$_2$ as discussed in supplementary material Sec.~S4. The analysis shows that the decrease of $\IC$ in the case of ML TMD/Py samples can be attributed to a negative sign of the spin-to-charge conversion. As a further confirmation of negative sign, we replace Py with Co$_{60}$Fe$_{20}$B$_{20}$ (CFB) [supplementary material Sec.~S4.] and show that negative spin to charge conversion is driven by spin-orbit coupling of the TMD layer.

In order to understand the observed behavior of $\jc^{TMD/Py}$ with respect to the type of TMD layer, we then calculated the 3D spin current density ($\js$) from the FMR measurements by using the theory of spin pumping~\cite{tserkovnyak2002enhanced}:

\begin{equation}\label{eq:js}
\begin{split}
\js=\frac{\smc\gamma^2 \hbar h_{\rm RF}^{2}}{8 \pi \alphaeff^{2}}\left[\frac{4 \pi \Ms \gamma + \sqrt{(4 \pi \Ms \gamma)^2+4\omega^2}}{(4 \pi \Ms \gamma)^2+4\omega^2}\right]\frac{2e}{\hbar},
\end{split}
\end{equation}

$h_{\rm RF}$ is the RF field generated due to the RF current of frequency $f=\omega/2\pi$ flowing through the co-planar waveguide (CPW), $\gamma$ is the gyromagnetic ratio, $4\pi \Ms$ is the saturation magnetization, $\alphaeff$ is the effective Gilbert damping constant, $\hbar$ is the reduced Planck's constant, and $e$ is the electronic charge. $\smc$ is the effective spin mixing conductance of the ML TMD-FM interface and denotes the efficiency of spin pumping. It can be calculated from measured damping constant as \cite{tserkovnyak2002enhanced}: 
\begin{figure*} [t!]
\includegraphics*[width=15.5cm]{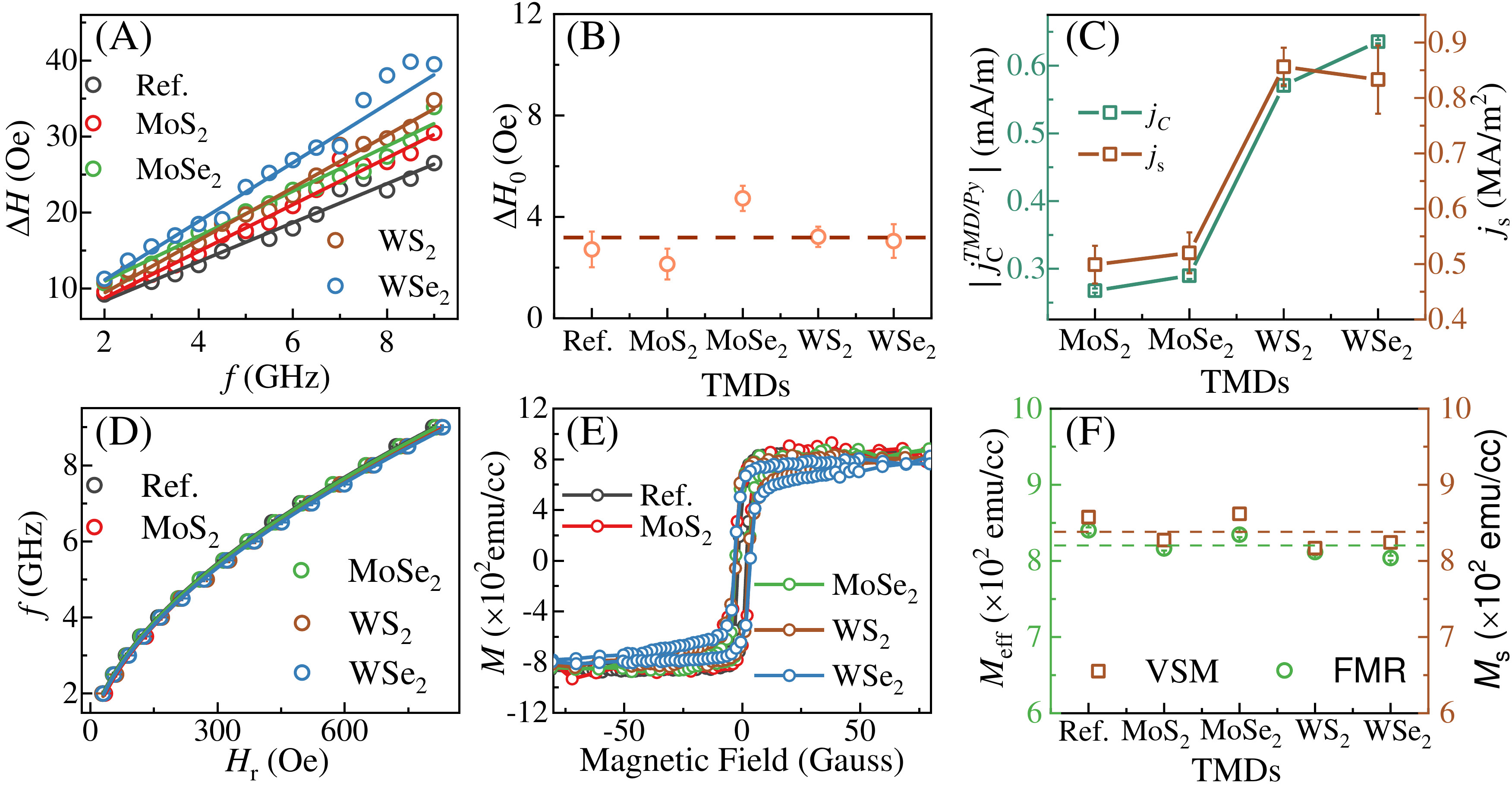}
\centering
\caption{\label{fig:4} \textbf{Determination of spin current density and magnetic properties} \textbf{(A)} The extracted data (symbols) for $\DelH$ versus $f$ and its fit to Eq.~\ref{eq;deltah} (solid line). \textbf{(B)}  The plot of $\Ho$ extracted from the fit in Fig.~4A. \textbf{(C)} Plot of $|\jc^{TMD/Py}|$ and $j_s$ for various ML TMDs, which follows the trend of SOC strength shown in Fig. 1E.  \textbf{(D)} $f$ vs. $\Hr$ data (points) fitted with Kittel's formula (solid lines) to extract $\Meff$. \textbf{(E)} Magnetization curves for various ML TMD/Py samples measured using VSM. \textbf{(F)} Comparative plot of $\Meff$ (green axis) and $\Ms$ values (brown axis) showing that the magnetic properties of Py is similar for all ML TMD/Py samples.}
\end{figure*}
\begin{equation}\label{eq:spinmix}
\begin{split}
 \smc=\frac{\Delta \alphaeff 4 \pi \Ms t_{Py}}{g\mu_B},
\end{split}
\end{equation}

where $t_{Py}$, $g$, $\mu_B$ denote Py layer thickness, the Land\'e $g$-factor, and the Bohr magneton, respectively. $\Delta \alphaeff=\alphaeff^{TMD/Py/Ta}-\alphaeff^{Py/Ta}$ represents the change in effective damping constant of the TMD/Py stacks with respect to the reference sample without the TMD layer. $\alphaeff$ was determined from the fitting of the linewidth ($\DelH$) versus frequency ($f$) data using the following equation:

\begin{equation}\label{eq;deltah}
\Delta H=\frac{2\pi\alphaeff f}{\gamma}+\Ho,
\end{equation}

where, $\Ho$ denotes the inhomogeneous line broadening. The slope of the linear fits (as shown in Fig.~4A) of the above equation gives $\alphaeff$, which is found to be higher for TMD/Py samples compared to the reference sample. $\alphaeff$  and hence $\Delta \alphaeff$ increase in the sequence MoS$_2\rightarrow$MoSe$_2\rightarrow$WS$_2\rightarrow$WSe$_2$ (Table~1). $\Ho$ is a measure of magnetic inhomogeneity and sample imperfections of the FM layer. The value of $\Ho$ for all the ML TMD/Py samples showed low values of $\Ho$ ($<$ 6 Oe), as shown in Fig.~4B, indicating that the Py quality is similar on the ML TMDs and comparable to the reference sample. We also determine the magnetization of Py to ascertain whether the magnetic characteristics of the Py layer formed on ML TMD layers have changed. We determine $\Meff$ from the $f$ versus $\Hr$ plot (Fig.~4D) by fitting with Kittel's equation~\cite{kittel1948theory}. The saturation magnetization $\Ms$  is measured using a vibrating-sample magnetometer (VSM) (Fig.~4E). $\Meff$ and $\Ms$ are found to be similar for all the ML TMDs/Py samples as shown in Fig.~4F, indicating that there is no significant difference in magnetic properties of the Py layers grown on the ML TMDs. Thus the decrease of $\IC$ with ML TMD can not be explained on the basis of a possible change of the magnetic properties of the Py layer. As illustrated in Fig.~4C, the $\js$ calculated using Eq.~\ref{eq:js} is found to increase in the sequence MoS$_2\rightarrow$MoSe$_2\rightarrow$WS$_2\rightarrow$WSe$_2$  which is similar to $\jc$. This indicates that the measured magnitude of the 2D charge current is directly proportional to the observed magnitude of the spin current.  

Our findings reveal that both $\js$ and $\jc$ increase in the sequence MoS$_2\rightarrow$MoSe$_2\rightarrow$WS$_2\rightarrow$WSe$_2$, the same sequence for which theoretically predicted SOC also increases (Fig.~1E). 
In the IREE mechanism, $\js$ is related to $\jc$ as~\cite{shen2014microscopic,sanchez2013spin}

\begin{equation}\label{lambda}
\begin{split}
\jc=\Lr \js,
\end{split}
\end{equation}

where, $\Lr=\alphaR\tau_{s}/\hbar$. Here 
$\tau_{s}$ is the momentum relaxation time~\cite{shen2014microscopic}. The $\Lr$ denotes the efficiency of spin-to-charge conversion. It has the unit of length as it is a ratio of 3D charge current density to the 2D spin current density (Eq.~5). The values of efficiency parameter $\Lr$ are summarized in Table~1. 
The value of $\Lr$ is found to increase in the sequence MoS$_2\rightarrow$MoSe$_2\rightarrow$WS$_2\rightarrow$WSe$_2$ and it lies in the range of $-0.54$ to $-0.76$~nm.  
Moreover, the values of $\Lr$ are significantly larger for W-based TMDs as compared to that of Mo-based TMDs. This may be due to stronger interfacial $d$-$d$ hybridization resulting from the higher spin-orbit coupling of W as compared to Mo atoms~\cite{tyer2003systematic}.

The magnitude of  $\Lr$ in our system is found to be significantly larger compared to other Rashba interfaces such as Bi/Ag~\cite{sanchez2013spin},  Ag/Sb~\cite{zhang2015spin}, Cu/Bi~\cite{isasa2016origin}, and NM/Bi$_2$O$_3$ (where NM denote non-magnetic metals Cu, Ag, Au, Al)~\cite{tsai2018clear} (see Fig. S8 for direct comparison). In the case of YIG/MoS$_2$ a value of $\Lr=0.4$~nm was reported~\cite{mendes2018efficient}. Our value is comparable to YIG/MoS$_2$ system. A recent theoretical work on ML TMD predicts the presence of a large orbital Hall effect (OHE) with a sign that is opposite to that of the SHE~\cite{bhowal2020intrinsic}. However, in our work, the FM layer is in direct contact with the TMD layer. Theoretical calculation for such ML TMD/Py interface is yet to be performed. The large and negative sign of spin-to-charge conversion in our work may appear to agree with the negative OHE predicted in Ref~\cite{bhowal2020intrinsic}. However, OHE is independent of SOC of the TMD layer, while in our experiment, we see an excellent correlation of $\Lr$ with SOC. Hence, we believe that the mechanism of spin-to-charge conversion in the present ML TMD/Py system is not due to the OHE, but rather it is governed by the SOC-dependent IREE.

For comparison of the spin-to-charge conversion efficiency of these ML TMDs with traditional 3D HMs, we also calculate the so-called pseudo spin Hall angle ($\TPSH$) as \cite{sanchez2013spin} 

\begin{equation}\label{PSH}
\begin{split}
\TPSH=2 \Lr/t
\end{split}
\end{equation}

where $t$ denote the thickness of ML TMDs. In the above equation, the thickness of the TMD is assumed to be less than the spin diffusion length~\cite{liang2017electrical}. Considering the thickness of Mo and W-based ML TMDs to be 0.7 and 0.8 nm, respectively~\cite{chang2014monolayer,zhang2013controlled} we estimated $\TPSH$ for all the ML TMDs as shown in Table~1. The  value of $\TPSH$ is found to be greater than 1 and at least one order higher than any of the reported values for heavy metals~\cite{zhang2015role,hoffmann2013spin,tiwari2017antidamping,kumar2021large,bansal2018large}. 
The enormous magnitude of pseudo spin Hall angle ($>$ 1) is unphysical for a 3D system, supporting our assertion that the spin-to-charge conversion process is driven by IREE rather than ISHE.
\section{CONCLUSIONS}

In summary, we demonstrate spin-to-charge conversion in large area monolayer TMDs, namely MoS$_2$, MoSe$_2$, WS$_2$ and WSe$_2$ with the efficiency that exceeds that of the traditional heavy metals such as Pt by about at least one order. We also demonstrate that the efficiency of spin-to-charge conversion is directly proportional to the SOC strength of the ML TMDs, indicating that the phenomenon is driven by the IREE effect arising from the two-dimensional interface of ML TMD and Py. Our findings suggest that the IREE has a negative sign of spin-to-charge conversion for the ML TMD/Py interface. The large spin-to-charge conversion of these interfaces is extremely promising for emerging spintronics applications such as for non-volatile memory and logic devices.  

\begin{center}

\begin{table*}
\begin{ruledtabular}
 \begin{tabular}{c c c c c c c} 
 
 \rowcolor{orange!50!white!40} \multicolumn{7}{c}{\textbf{Table 1:} Values of effective damping parameter subtracted from reference sample} \\
 
 \rowcolor{orange!50!white!40} \multicolumn{7}{c}{($\Delta \alphaeff$), spin mixing conductance ($\smc $), charge current density ($\jc $), spin current} \\
  \rowcolor{orange!50!white!40} \multicolumn{7}{c}{density ($\js $), Rashba efficiency parameter ($\Lr$), pseudo spin Hall angle ($\TPSH$)} \\
 
 TMD & $\Delta \alphaeff$ & $\smc $ & $\jc $ & $\js $ & $\Lr $ & $\TPSH$ \\ [0.5ex] 
 
   & $(\times 10^{-3})$ & $ (\times 10^{19} \textrm{m}^{-2}) $ & $(\mu \textrm{A/m})$ & $(\times 10^{4}\textrm{A}/\textrm{m}^{2})$ & $(\times 10^{-10} \textrm{m})$ &  \\ [0.5ex] 
 \hline

 MoS$_2$ & 1.4 $\pm$ 0.1 & 0.72 $\pm$ 0.04 & $-267\pm$ 3 & 50 $\pm$ 3 & -5.4 $\pm$ 0.4 & -1.5 $\pm$ 0.1 \\
 
 MoSe$_2$ & 1.5 $\pm$ 0.1 & 0.78 $\pm$ 0.05 & $-290\pm$ 5 & 52 $\pm$ 4 & -5.6 $\pm$ 0.4 & -1.6 $\pm$ 0.1 \\

 WS$_2$ & 2.4 $\pm$ 0.1 & 1.26 $\pm$ 0.04 & $-571\pm$ 2 & 86 $\pm$ 3 & -6.7 $\pm$ 0.3 & -1.7 $\pm$ 0.1 \\
 
 WSe$_2$ & 3.6 $\pm$ 0.2 & 1.84 $\pm$ 0.11 & $-635\pm$ 2 & 83 $\pm$ 6 & -7.6 $\pm$ 0.6 & -1.9 $\pm$ 0.2 \\ [1ex] 
 
\end{tabular}
\end{ruledtabular}
\end{table*}
\end{center}

\section{METHODS}

\subsection*{Material growth}
We prepared Al$_2$O$_3$/ML TMD/Py(10 nm)/Ta(3 nm) for four different ML TMDs, namely MoS$_2$, MoSe$_2$, WS$_2$ and WSe$_2$ along with a reference sample Al$_2$O$_3$/Py(10 nm)/Ta(3 nm) simultaneously. Py thin film of 10 nm thickness was grown using AJA ATC Orion 8 ultra-high vacuum magnetron sputtering system on commercially purchased chemical vapor deposited monolayer TMD samples (from 2D semiconductors Inc.) on the $c$-cut sapphire substrates (with the size of $5\times2$ mm$^2$). The base pressure of the sputtering chamber was better than 8$\times 10^{-8}$\, Torr
and the working pressure was 2$\times 10^{-3}$\, Torr. A Tantalum (Ta) layer of 3 nm thickness was deposited as a capping layer to prevent oxidation of Py thin film. 
\subsection*{Raman measurements}
Raman spectroscopy was used to determine the thickness and quality of TMD samples before and after Py layer deposition. The measurements were performed using Renishaw inVia confocal microscope with 532 nm laser wavelength and 2400 lines per mm grating at 5 mW power. 

\subsection*{Magnetostatic characterization}
The saturation magnetization of the reference sample Py(10 nm)/Ta(3 nm) and ML TMD/Py(10 nm)/Ta(3 nm) samples were determined using the VSM module of  physical property measurement system (PPMS) manufactured by Quantum Design  (Model-PPMS EverCool-II ). All the VSM measurements were performed at room temperature.

\subsection*{Spin pumping and IREE characterization}

Broadband FMR spectroscopy technique was used to characterize the magnetization dynamics. The sample was placed upside down on the co-planar waveguide (CPW), through which RF current is passed, generating an RF field to excite the resonance in the Py layer. Field modulation technique with lock-in-based detection is employed to enhance the sensitivity of FMR measurements. For this purpose, the external DC field was modulated with a low frequency (213 Hz) AC field generated in the Helmholtz coils using the lock-in amplifier. The RF excitation frequency was varied from 2 to 9 GHz. The spectrum obtained by field sweep is then fitted using the derivative of the Lorentzian line-shape function to extract ($\Hr$) and ($\DelH$). $h_{ \rm RF}$ is calculated to be 0.35 G using Amp\`ere's law, $h_{\rm RF} = \frac{\mu_{\rm 0}}{2\pi w} \sqrt[]{\frac{P_{\rm RF}}{Z}}\log(1+\frac{w}{D})$, where $Z$ is the impedance of CPW (50~$\Omega$), $P_{\rm RF}$ is the applied RF power, $w$ is the width of the signal line (0.25 mm), and $D$ is the separation between the signal line to the sample (0.5 mm).

For the IREE measurements, electrical contacts were made using copper pads beneath the inverted sample. We employed a field modulation technique, which involves modulating the DC field with a small AC field created by a pair of Helmholtz coils.~\cite{kumar2018large} The coils are powered by a lock-in amplifier which also measures the voltage generated by the IREE. For IREE measurements, RF microwave power was set at 0.5 W and the RF frequency was set at 3 GHz.

\section{acknowledgement}
The partial support from the Ministry of Human Resource Development under the IMPRINT program (Grant no: 7519 and 7058), the Department of Electronics and Information Technology (DeitY), 
Science \& Engineering research board (SERB File no. CRG/2018/001012), Joint Advanced
Technology Centre at IIT Delhi, Grand Challenge Project, IIT Delhi, and Department of Science and Technology under the Nanomission program (grant no: $SR/NM/NT-1041/2016(G)$) are gratefully acknowledged. H.B. gratefully acknowledges the financial support from the Council of Scientific and Industrial Research (CSIR), Government of India. 




\section{Supplementary information}

 \subsection{Raman measurements}

\begin{figure*} [t!]
\includegraphics*[width=15.5cm]{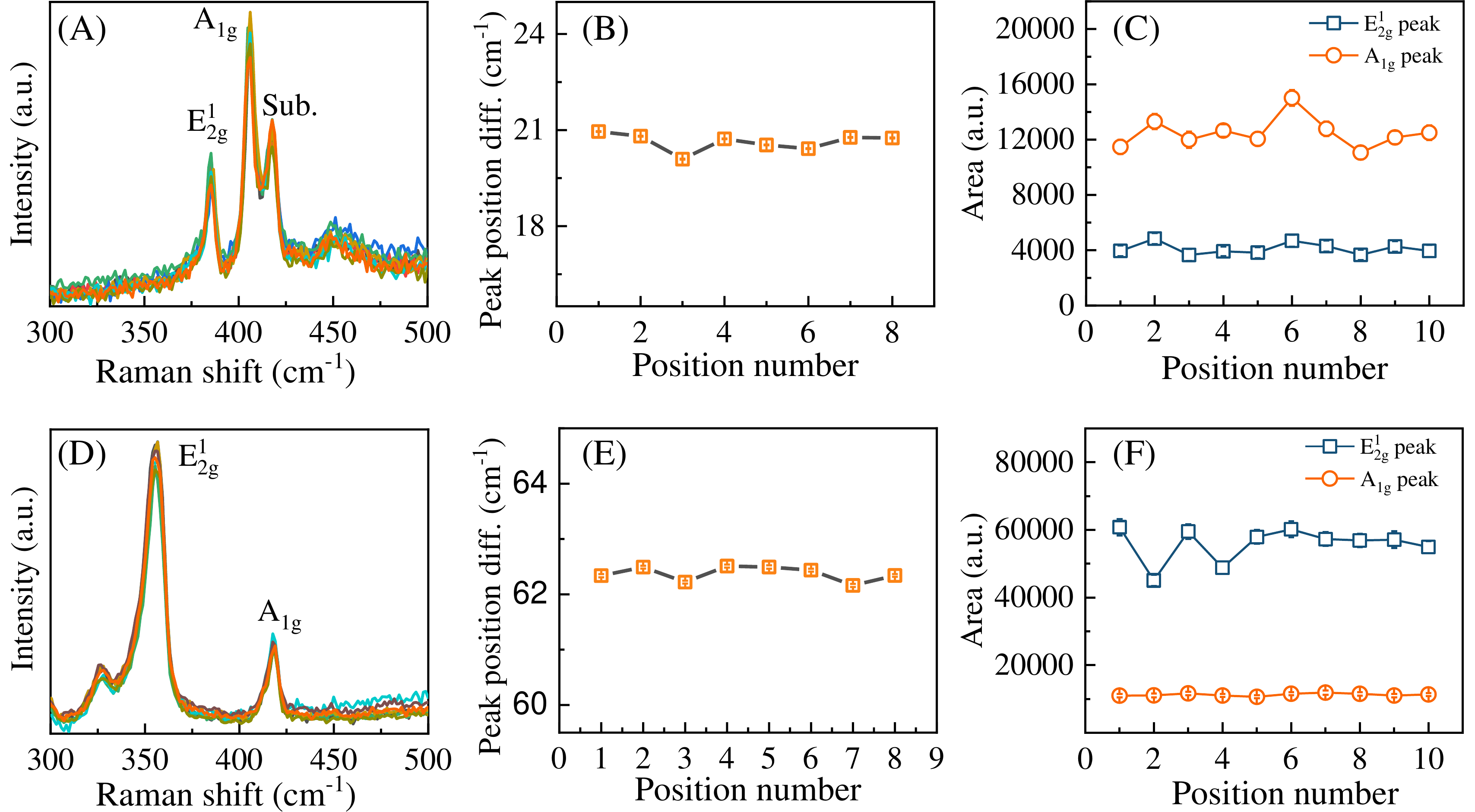}
\caption{\label{fig:1_1} \textbf{Raman measurements.} \textbf{(A)} Raman spectra at 8 different points for \textbf{(A)} MoS$_2$ and \textbf{(D)} WS$_2$. The peak position difference between A$_{\rm1g}$ and E$_{\rm2g}^{1}$ peaks for \textbf{(B)} MoS$_2$ and \textbf{(E)} WS$_2$. Trend of intensity for \textbf{(C)} MoS$_2$ and \textbf{(F)} WS$_2$ to confirm uniformity of the TMDs.}
\end{figure*}

\begin{figure*} [t!]
\includegraphics*[width=10cm]{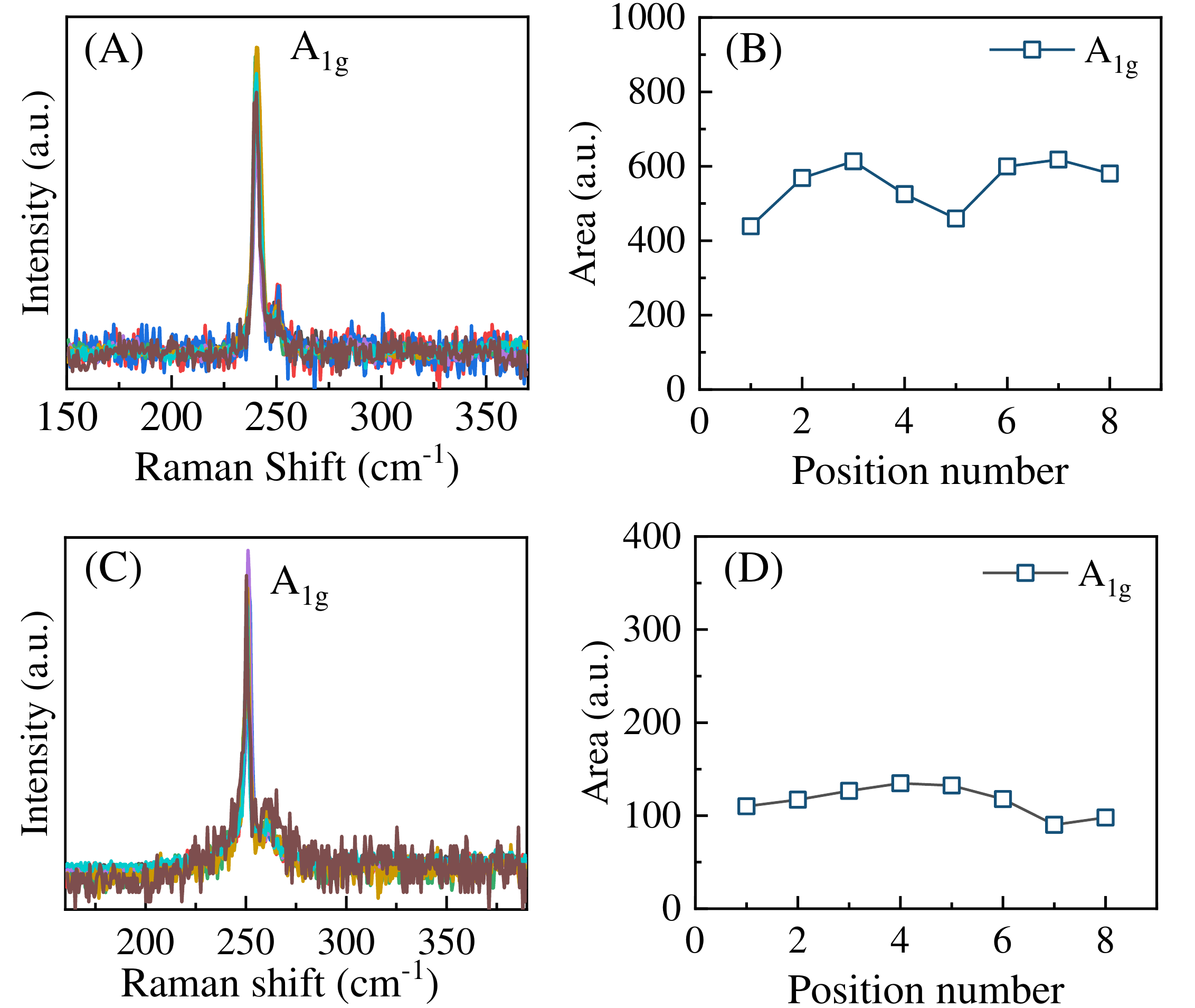}
\caption{\label{fig:1_2} \textbf{Raman measurements.} \textbf{(A)} Raman spectra at 8 different points for \textbf{(A)} MoSe$_2$ and \textbf{(C)} WSe$_2$. Trend of intensity for \textbf{(B)} MoSe$_2$ and \textbf{(D)} WSe$_2$ to confirm uniformity of the TMDs.}
\end{figure*}

To validate that the commercial CVD grown-monolayer (ML) TMDs were completely covered, we performed Raman measurements at several locations on the sample using the same measurement conditions (Laser wavelength of 532 nm, 2400 lines per mm grating at 5 mW power). The laser spot size was approximately 0.8~$\mu$m. The location was chosen at random to encompass the whole $5\times2$ mm$^2$ sample region. The Raman data for MoS$_2$ are shown in Fig.~\ref{fig:1_1} (A), while the difference in peak position between the E$_{\rm 2g}^{1}$ and A$_{\rm 1g}$ peaks and the intensity of the peaks is shown in Fig.~\ref{fig:1_1} (B) and (C) respectively, for several locations on the sample. The variation in peak location and the intensity is essentially identical for all locations, confirming the presence of large-area ML MoS$_2$ over the entire substrate. A similar result is obtained for WS$_2$ as shown in Fig. \ref{fig:1_1} (D), (E) and (F). The absence of the B$_{\rm2g}^{1}$ mode for MoSe$_2$ and WSe$_2$ is employed as a signature to determine the ML thickness. The position of B$_{\rm2g}^{1}$ mode for MoSe$_2$ and WSe$_2$ are 353~cm$^{-1}$ and 309~cm$^{-1}$, respectively.~\cite{tonndorf2013photoluminescence} As demonstrated in Fig.~\ref{fig:1_2} (A) and Fig.~S2 (C), these peaks are absent from all the measured locations in both MoSe$_2$ and WSe$_2$. The trend of the peak intensity for MoSe$_2$ and WSe$_2$ is shown in Fig.~\ref{fig:1_2} (B) and Fig.~\ref{fig:1_2} (D) respectively, which is nearly constant. As a result, we conclude that all of our TMD samples cover the entire area of the sample. We observed a weak defect peak near the main A$_{\rm1g}$ peak in MoSe$_2$ and WSe$_2$ samples. This peak appears due to Se vacancies, and it can be utilized to estimate the concentration of Se vacancies in the sample.~\cite{mahjouri2016tailoring}  Based on the A$_{\rm1g}$/(defect peak) ratio [Fig.~\ref{fig:1_2} (A) and (C)], we estimate the selenium (Se) vacancies in both the MoSe$_2$ and WSe$_2$ to be $<2.5$\%~\cite{mahjouri2016tailoring}.

\subsection{Surface Morphology by Atomic Force Microscopy}

\begin{figure*} [ht]
\begin{center}
\includegraphics*[width=14.5cm]{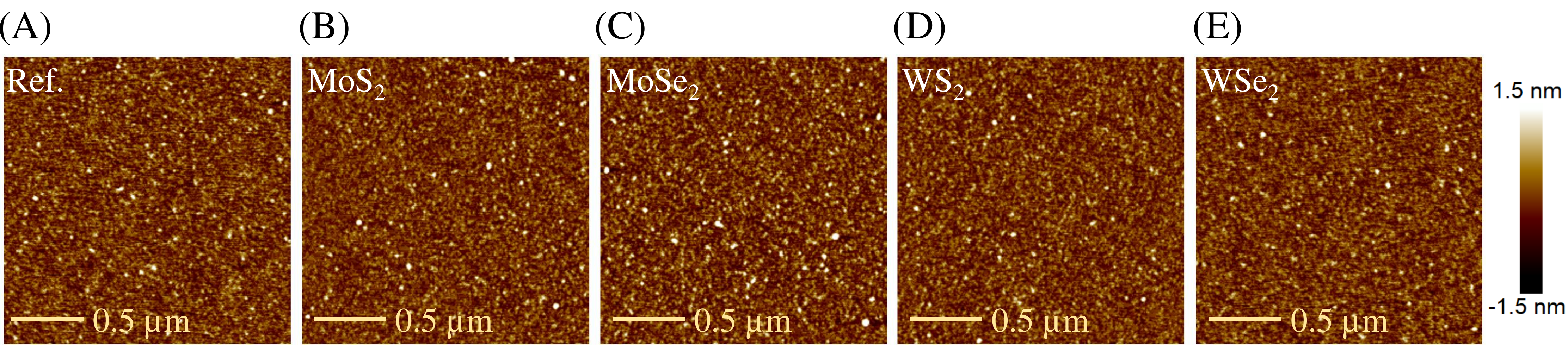}
\caption{\label{fig:5} \textbf{ Surface morphology.} Atomic force microscopy (AFM) images of \textbf{(A)} reference, \textbf{(B)} MoS$_2$, \textbf{(C)} MoSe$_2$, \textbf{(D)} WS$_2$, \textbf{(E)} WSe$_2$ samples respectively. All the samples have similar morphology and roughness as they are prepared simultaneously inside sputtering chamber.}
\end{center}
\end{figure*}

Our 2D TMD ML have atomic roughness ($<0.15$~nm). Hence we expect good quality of thin films grown on top the 2D TMD ML samples. We performed atomic force microscopy (AFM) measurements to study the surface morphology and to determine the surface roughness. The measurements were performed in tapping mode using Bruker AFM system. Figure~\ref{fig:5} shows the AFM images for the (A) reference, (B) MoS$_2$, (C) MoSe$_2$, (D) WS$_2$, (E) WSe$_2$ samples. Surface morphology of all the samples is similar as they were prepared simultaneously inside the sputtering chamber. The root mean square (rms) surface roughness for reference, MoS$_2$, MoSe$_2$, WS$_2$, and WSe$_2$ samples are 0.32, 0.30, 0.34, 0.29, and 0.30 nm respectively. This indicates high quality of the films.

\begin{figure} [b]
\begin{center}
\includegraphics*[width=6cm]{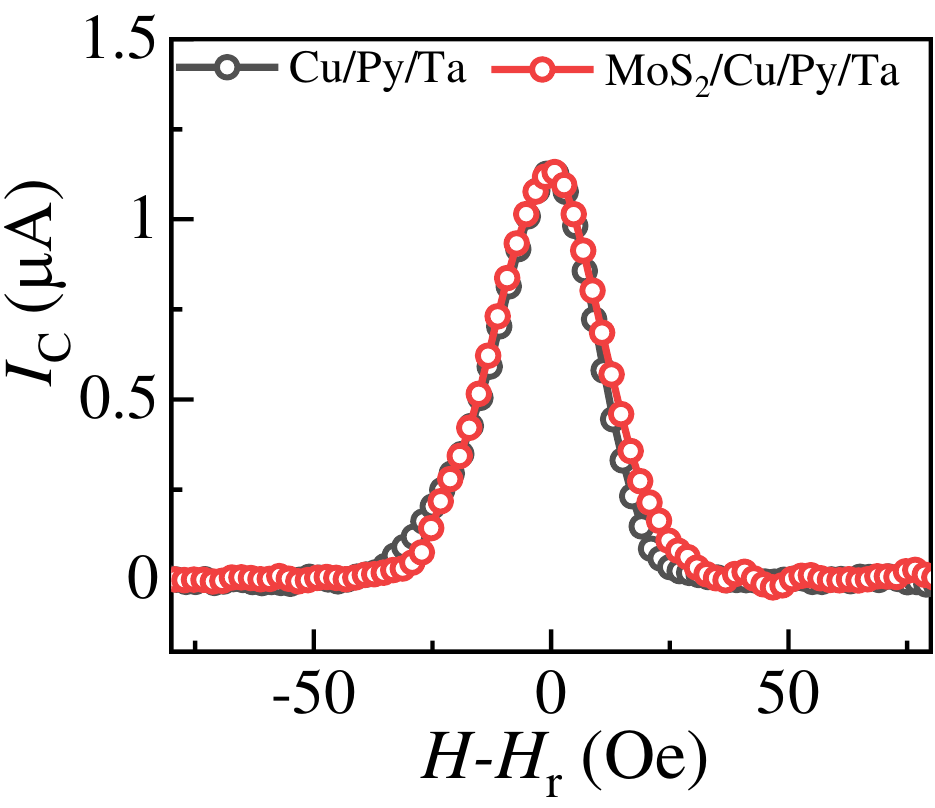}
\caption{\label{fig:S2} \textbf{Effect of Cu spacer.} Measured $\IC$ for ML MoS$_2$/Cu(5 nm)/Py(10 nm)/Ta(3 nm) (Red symbols) and  Cu(5 nm)/Py(10 nm)/Ta(3 nm) (black symbols).}
\end{center}
\end{figure}
\subsection{Effect of copper spacer}

\begin{figure*} [ht]
\begin{center}
\includegraphics*[width=15.5cm]{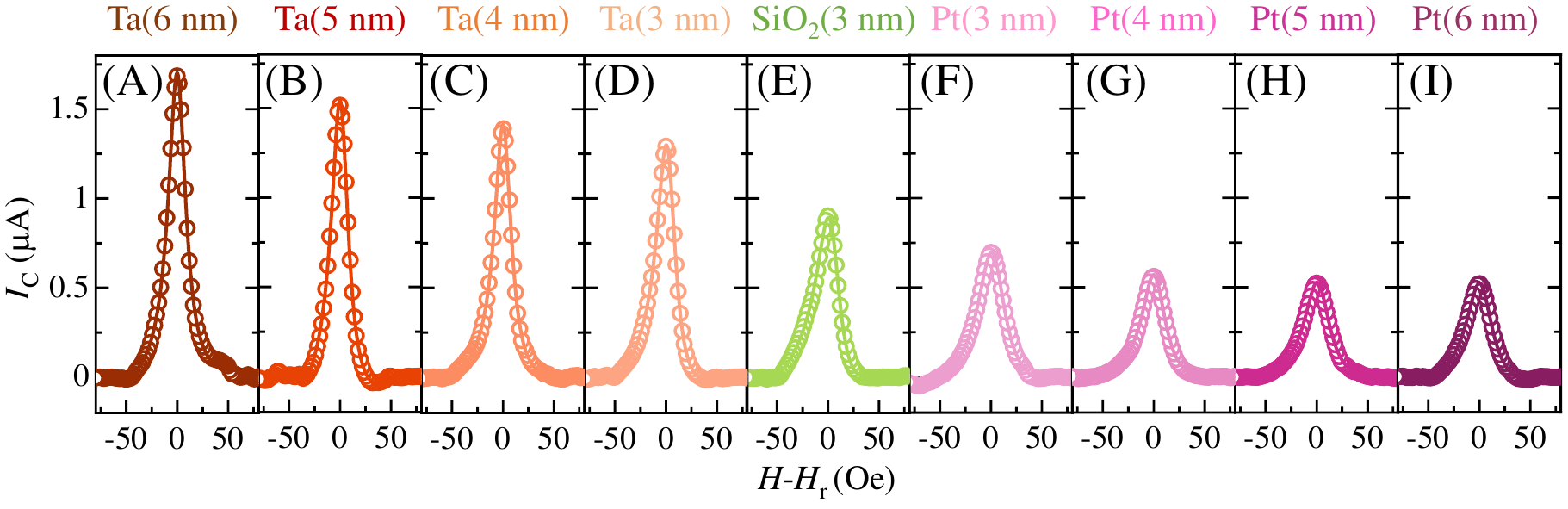}
\caption{\label{fig:S3} \textbf{Effect of capping layer on the magnitude of $\IC$.} Plot of $\IC$ for Py (10 nm) samples having different capping layers, namely, Pt, Ta, and SiO$_2$.}
\end{center}
\end{figure*}
To confirm that the decrease of $\IC$ in TMD/Py/Ta samples was driven by IREE from TMD/Py interface, we grew samples with a Cu spacer layer: TMD/Cu(5 nm)/Py(10 nm)/Ta(3 nm). The Cu layer is used to break the interface between the TMDs and Py layer. Figure~\ref{fig:S2} shows the IREE signal for the sample with and without Cu insertion layer for MoS$_2$. The reference sample without the TMD layer was grown in the same chamber together with the TMD/Cu/Py sample so that measurement artifacts due to the variation in growth conditions can be removed. The reference sample (black circles in Fig. \ref{fig:S2}) and the TMD/Cu/Py have the same magnitude of $\IC$, as illustrated in the figure. The Cu layer has a spin diffusion length, which is larger than the thickness of Cu (5 nm).~\cite{villamor2013contribution,yakata2006temperature} Thus, the ISHE signals due to the Py layer itself and the Ta capping layer (which are bulk effects) are unaffected by the presence of the Cu layer. The IREE signal from the TMD/Py interface, on the other hand, will be affected. We detect no decrease of $\IC$ in the figure, in contrast to the scenario when the TMD layer is in direct contact with the Py layer [Fig. 3 of the main text]. As a result, the decrease of $\IC$ seen in the main text is interfacial in nature and can be attributed to IREE.

\subsection{Determining the sign of IREE signal}

In the main text, we observe a decrease of $\IC$ when the TMD layer directly contacts the Py layer. We attribute this behavior to a negative spin-to-charge conversion. To further corroborate the sign of spin-to-charge conversion, we prepared a series of sample stack of Al$_2$O$_3$/Py (10 nm) with Pt and Ta capping layers, which are known to exhibit the opposite sign of spin-to-charge conversion.  Figure~\ref{fig:S3} shows measured $\IC$ for the capping layer of Ta, Pt, and SiO$_2$. We observe a finite signal even in the case of Py/SiO$_2$, the origin of which is discussed in the following section. We observe a decrease of $\IC$ with Pt capping and an increase in $\IC$ with Ta capping-consistent with the opposite sign of spin-to-charge conversion for Pt and Ta. The sign of spin-to-charge conversion also depends on the stacking order of the FM and HM layers. In Fig. 3 of the main text, the TMDs layers are beneath the Py layer, whereas in Fig.~\ref{fig:S3}, the Pt or Ta layer is above the Py layer. Hence, the decrease of $\IC$ for TMD/Py samples relative to the corresponding reference sample implies that the sign of spin to charge conversion at the TMD/Py interface is similar to that of Ta, which has negative spin-to-charge conversion relative to Pt.~\cite{kumar2018large,kumar2021large,yu2018determination} Thus we conclude that the drop in $\IC$ with the TMD layer is owing to a negative sign of spin-to-charge conversion at the interface of TMD and Py layer. This is further supported by scaling of $\IC^{TMD/Py}$ ($\IC$ from the interface of TMD and Py) with the SOC of the TMD layer.

\begin{figure} [t!]
\begin{center}
\includegraphics*[width=6cm]{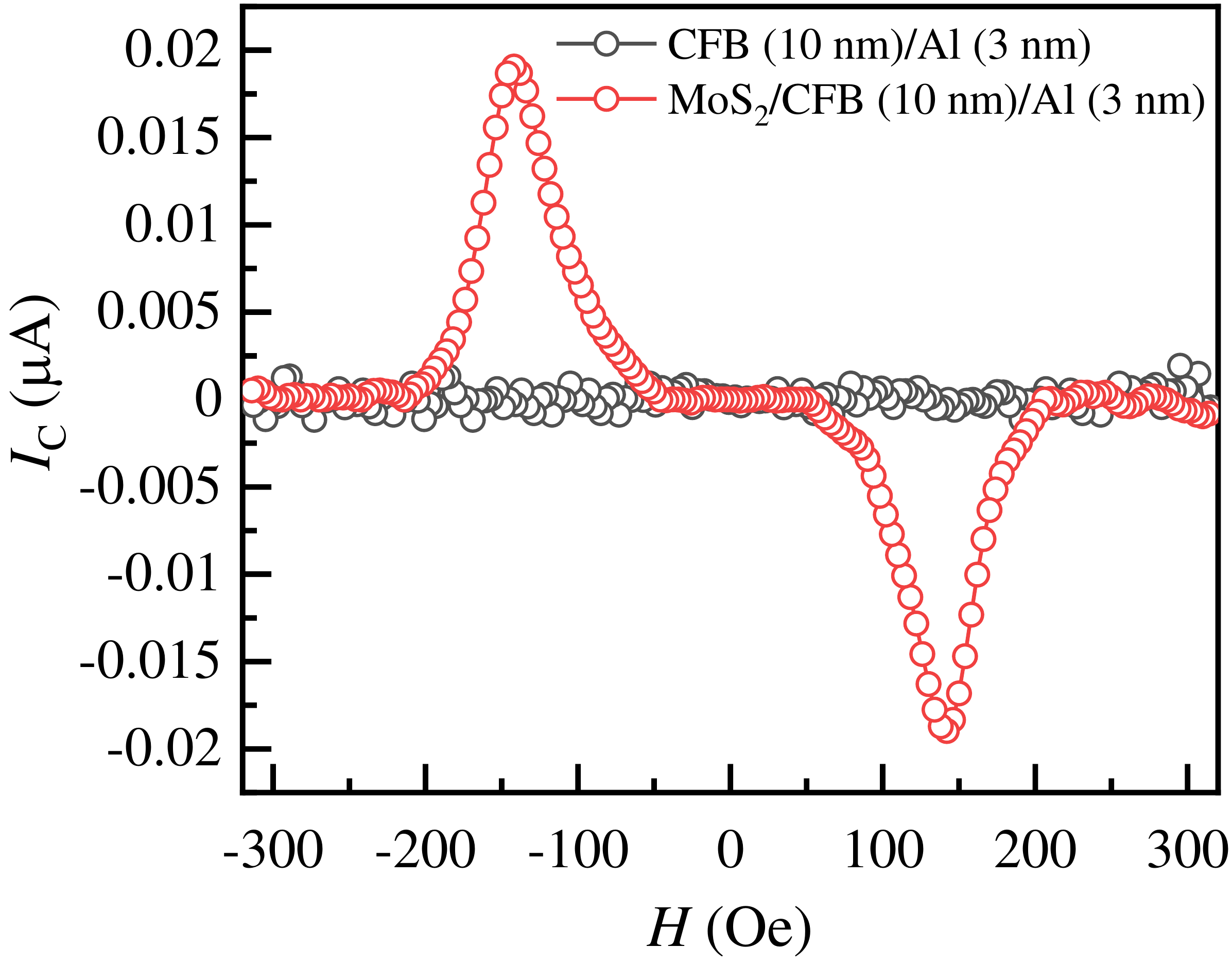}
\caption{\label{fig:6} Measured $\IC$ for CFB (10 nm)/Al (3 nm) (black) and MoS$_2$ (ML)/CFB (10 nm)/Al (3 nm) (red). The measurement conditions were kept identical with Fig.~\ref{fig:3}.}
\end{center}
\end{figure}

To further confirm the negative sign of $\IC^{TMD/Py}$, we also replace Py with Co$_{60}$Fe$_{20}$B$_{20}$ (CFB) for which the self-induced ISHE is known to be negligible.~\cite{gladii2019self} We simultaneously prepared two samples: CFB (10 nm)/Al (3 nm) and MoS$_2$ (ML)/CFB (10 nm)/Al (3 nm). As shown in Fig.~\ref{fig:6}, we did not observe any signal from the CFB (10 nm)/Al (3 nm) sample showing the absence of self-induced ISHE signal in CFB sample. In contrast, the MoS$_2$ (ML)/CFB (10 nm)/Al (3 nm) sample produce a clear negative signal (for same measurement conditions as in Fig.~\ref{fig:3}) establishing a negative spin to charge conversion driven by spin-orbit coupling in MoS$_2$ rather than the FM layer. Please note that in Fig.~\ref{fig:3} all measurements are shown for positive magnetic field. Furthermore, in Fig.~\ref{fig:3} the heavy metal Pt and Ta are capping layers. Hence the stacking order need to be taken in to the account while comparing sign of the signal between Fig.~\ref{fig:3} and Fig.~\ref{fig:6}.

\begin{figure} [ht!]
\begin{center}
\includegraphics*[width=6cm]{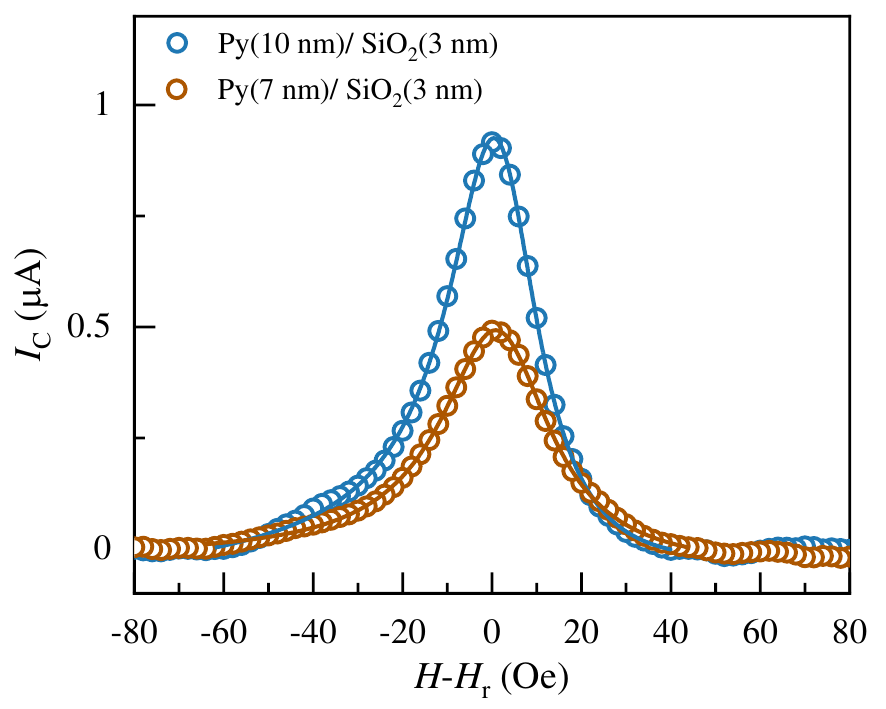}
\caption{\label{fig:S4} \textbf{ Self-induced ISHE in Py/SiO$_2$.} Measured $\IC$ in Py/SiO$_2$ for two different thicknesses of the Py layer.}
\end{center}
\end{figure}
\begin{figure} [ht!]
\includegraphics*[width=7cm]{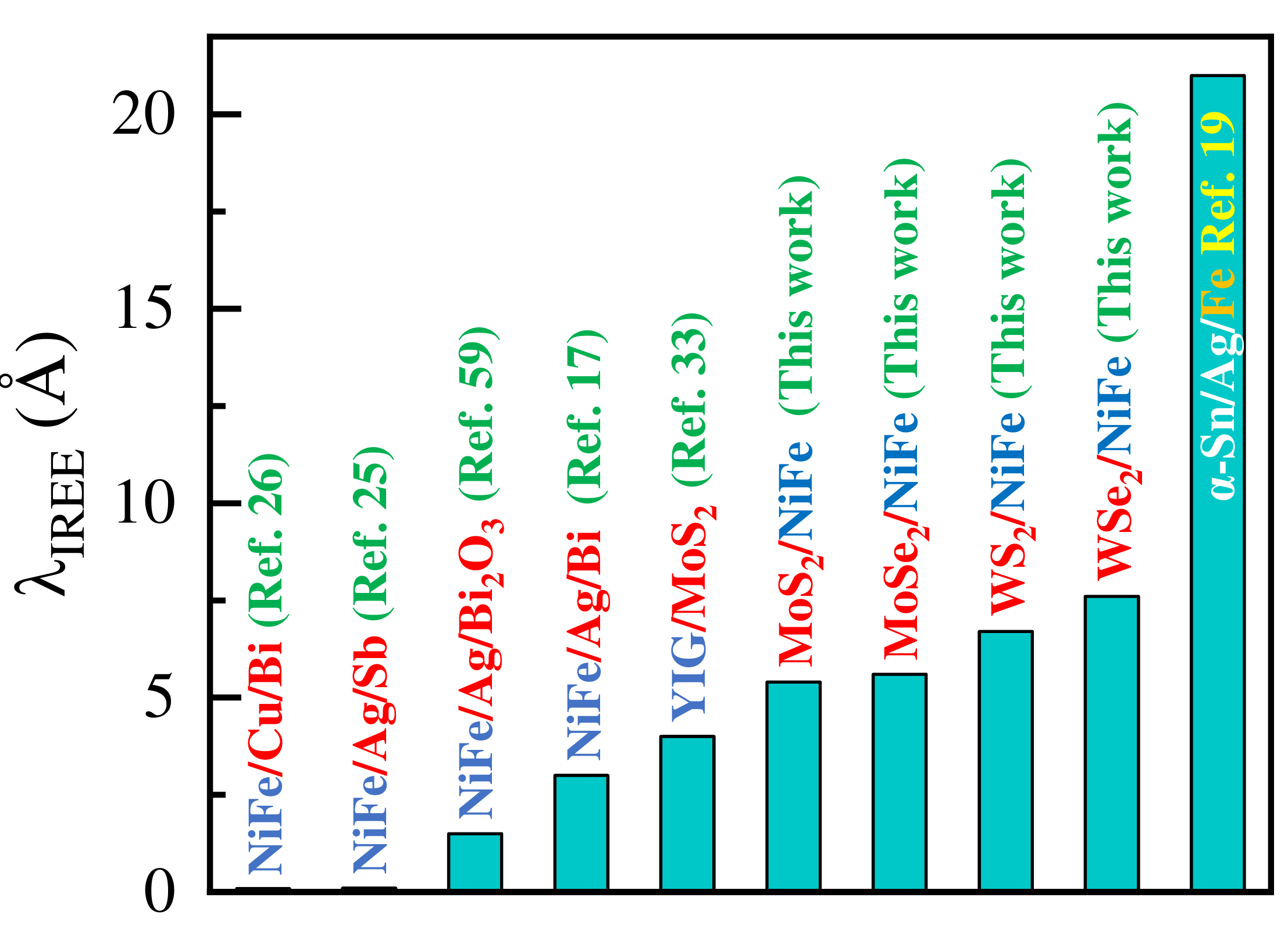}
\caption{\label{fig:8} Comparative plot of $\Lr$ for different Rashba interfaces.}
\end{figure}
\newpage
\subsection{Origin of \texorpdfstring{$\IC$}{tau2} in Py/SiO\texorpdfstring{$_2$}{tau1}}

We observe a finite $\IC$ in Py/SiO$_2$, which contains no HM layer. Additionally, the line shape is almost symmetrical with SiO$_2$ capping, which indicates that the signal is not due to galvanic effects.~\cite{lustikova2015vector} We believe that the signal arises from self-induced inverse spin Hall effect (ISHE), which has been reported recently for Py layer via the spin-pumping-induced spin current ISHE measurement.~\cite{tsukahara2014self,gladii2019self} Because the ISHE is a bulk effect, the self-induced ISHE signal should increase with the increase in thickness of the Py layer. In Fig.~\ref{fig:S4}, we show measured $\IC$ in Py/SiO$_2$ for two examples with different thicknesses of the Py layer. We see a clear increase of signal with the thickness of Py, which is a signature of self-induced ISHE.~\cite{gladii2019self} Relatively larger self-induced ISHE signal observed in our work is due to larger thickness of Py used in our work, which was chosen to obtain better signal strength in ISHE measurements.

\subsection{Comparison of \texorpdfstring{$\Lr$}{tau3} with other Rashba interfaces}

In Fig. S8, we compare the  magnitude of  $\Lr$ in our study with selected Rashba interfaces reported in literature. We found our $\Lr$ to be significantly larger compared to other Rashba interfaces such as Cu/Bi~\cite{isasa2016origin}, Ag/Sb~\cite{zhang2015spin}, NM/Bi$_2$O$_3$ (where NM denote non-magnetic metals Cu, Ag, Au, Al)~\cite{tsai2018clear}, Bi/Ag~\cite{sanchez2013spin}, and YIG/MoS$_2$~\cite{mendes2018efficient} with exception of  $\alpha$-Sn/Ag~\cite{rojas2016spin}, for which a larger value is reported.

\newpage
\bibliography{Reference.bib}

\providecommand{\latin}[1]{#1}
\makeatletter
\providecommand{\doi}
  {\begingroup\let\do\@makeother\dospecials
  \catcode`\{=1 \catcode`\}=2 \doi@aux}
\providecommand{\doi@aux}[1]{\endgroup\texttt{#1}}
\makeatother
\providecommand*\mcitethebibliography{\thebibliography}
\csname @ifundefined\endcsname{endmcitethebibliography}
  {\let\endmcitethebibliography\endthebibliography}{}
\begin{mcitethebibliography}{70}
\providecommand*\natexlab[1]{#1}
\providecommand*\mciteSetBstSublistMode[1]{}
\providecommand*\mciteSetBstMaxWidthForm[2]{}
\providecommand*\mciteBstWouldAddEndPuncttrue
  {\def\EndOfBibitem{\unskip.}}
\providecommand*\mciteBstWouldAddEndPunctfalse
  {\let\EndOfBibitem\relax}
\providecommand*\mciteSetBstMidEndSepPunct[3]{}
\providecommand*\mciteSetBstSublistLabelBeginEnd[3]{}
\providecommand*\EndOfBibitem{}
\mciteSetBstSublistMode{f}
\mciteSetBstMaxWidthForm{subitem}{(\alph{mcitesubitemcount})}
\mciteSetBstSublistLabelBeginEnd
  {\mcitemaxwidthsubitemform\space}
  {\relax}
  {\relax}

\bibitem[Liu \latin{et~al.}(2019)Liu, Lam, Zhu, Zheng, Li, Du, Liu, and
  Pong]{liu2019overview}
Liu,~X.; Lam,~K.; Zhu,~K.; Zheng,~C.; Li,~X.; Du,~Y.; Liu,~C.; Pong,~P.~W.
  \emph{IEEE Trans. Magn.} \textbf{2019}, \emph{55}, 1--22\relax
\mciteBstWouldAddEndPuncttrue
\mciteSetBstMidEndSepPunct{\mcitedefaultmidpunct}
{\mcitedefaultendpunct}{\mcitedefaultseppunct}\relax
\EndOfBibitem
\bibitem[Jung \latin{et~al.}(2022)Jung, Lee, Myung, Kim, Yoon, Kwon, Ju, Kim,
  Yi, Han, Kwon, Seo, Lee, Koh, Lee, Song, Choi, Ham, and
  Kim]{jung2022crossbar}
Jung,~S. \latin{et~al.}  \emph{Nature} \textbf{2022}, \emph{601},
  211--216\relax
\mciteBstWouldAddEndPuncttrue
\mciteSetBstMidEndSepPunct{\mcitedefaultmidpunct}
{\mcitedefaultendpunct}{\mcitedefaultseppunct}\relax
\EndOfBibitem
\bibitem[Gambardella and Miron(2011)Gambardella, and
  Miron]{gambardella2011current}
Gambardella,~P.; Miron,~I.~M. \emph{Phil. Trans. R. Soc. A.} \textbf{2011},
  \emph{369}, 3175--3197\relax
\mciteBstWouldAddEndPuncttrue
\mciteSetBstMidEndSepPunct{\mcitedefaultmidpunct}
{\mcitedefaultendpunct}{\mcitedefaultseppunct}\relax
\EndOfBibitem
\bibitem[Liu \latin{et~al.}(2012)Liu, Pai, Li, Tseng, Ralph, and
  Buhrman]{liu2012spin}
Liu,~L.; Pai,~C.-F.; Li,~Y.; Tseng,~H.; Ralph,~D.; Buhrman,~R. \emph{Science}
  \textbf{2012}, \emph{336}, 555--558\relax
\mciteBstWouldAddEndPuncttrue
\mciteSetBstMidEndSepPunct{\mcitedefaultmidpunct}
{\mcitedefaultendpunct}{\mcitedefaultseppunct}\relax
\EndOfBibitem
\bibitem[Manchon \latin{et~al.}(2019)Manchon, {\v{Z}}elezn{\`y}, Miron,
  Jungwirth, Sinova, Thiaville, Garello, and Gambardella]{Manchon2019review}
Manchon,~A.; {\v{Z}}elezn{\`y},~J.; Miron,~I.~M.; Jungwirth,~T.; Sinova,~J.;
  Thiaville,~A.; Garello,~K.; Gambardella,~P. \emph{Rev. Mod. Phys.}
  \textbf{2019}, \emph{91}, 035004\relax
\mciteBstWouldAddEndPuncttrue
\mciteSetBstMidEndSepPunct{\mcitedefaultmidpunct}
{\mcitedefaultendpunct}{\mcitedefaultseppunct}\relax
\EndOfBibitem
\bibitem[Kumar \latin{et~al.}(2021)Kumar, Sharma, Ali~Khan, Murapaka, Lim, Lew,
  Chaudhary, and Muduli]{kumar2021large}
Kumar,~A.; Sharma,~R.; Ali~Khan,~K.~I.; Murapaka,~C.; Lim,~G.~J.; Lew,~W.~S.;
  Chaudhary,~S.; Muduli,~P.~K. \emph{ACS Appl. Electron. Mater.} \textbf{2021},
  \emph{3}, 3139--3146\relax
\mciteBstWouldAddEndPuncttrue
\mciteSetBstMidEndSepPunct{\mcitedefaultmidpunct}
{\mcitedefaultendpunct}{\mcitedefaultseppunct}\relax
\EndOfBibitem
\bibitem[Sinova \latin{et~al.}(2015)Sinova, Valenzuela, Wunderlich, Back, and
  Jungwirth]{sinova2015spin}
Sinova,~J.; Valenzuela,~S.~O.; Wunderlich,~J.; Back,~C.~H.; Jungwirth,~T.
  \emph{Rev. Mod. Phys.} \textbf{2015}, \emph{87}, 1213--1260\relax
\mciteBstWouldAddEndPuncttrue
\mciteSetBstMidEndSepPunct{\mcitedefaultmidpunct}
{\mcitedefaultendpunct}{\mcitedefaultseppunct}\relax
\EndOfBibitem
\bibitem[Dyakonov and Perel(1971)Dyakonov, and Perel]{dyakonov1971current}
Dyakonov,~M.~I.; Perel,~V. \emph{Phys. Lett. A} \textbf{1971}, \emph{35},
  459--460\relax
\mciteBstWouldAddEndPuncttrue
\mciteSetBstMidEndSepPunct{\mcitedefaultmidpunct}
{\mcitedefaultendpunct}{\mcitedefaultseppunct}\relax
\EndOfBibitem
\bibitem[Hirsch(1999)]{hirsch1999spin}
Hirsch,~J.~E. \emph{Phys. Rev. Lett.} \textbf{1999}, \emph{83},
  1834--1837\relax
\mciteBstWouldAddEndPuncttrue
\mciteSetBstMidEndSepPunct{\mcitedefaultmidpunct}
{\mcitedefaultendpunct}{\mcitedefaultseppunct}\relax
\EndOfBibitem
\bibitem[Khvalkovskiy \latin{et~al.}(2013)Khvalkovskiy, Cros, Apalkov, Nikitin,
  Krounbi, Zvezdin, Anane, Grollier, and Fert]{khvalkovskiy2013matching}
Khvalkovskiy,~A.; Cros,~V.; Apalkov,~D.; Nikitin,~V.; Krounbi,~M.; Zvezdin,~K.;
  Anane,~A.; Grollier,~J.; Fert,~A. \emph{Phys. Rev. B} \textbf{2013},
  \emph{87}, 020402\relax
\mciteBstWouldAddEndPuncttrue
\mciteSetBstMidEndSepPunct{\mcitedefaultmidpunct}
{\mcitedefaultendpunct}{\mcitedefaultseppunct}\relax
\EndOfBibitem
\bibitem[Chen \latin{et~al.}(2016)Chen, Dumas, Eklund, Muduli, Houshang, Awad,
  D{\"u}rrenfeld, Malm, Rusu, and {\AA}kerman]{chen2016spin}
Chen,~T.; Dumas,~R.~K.; Eklund,~A.; Muduli,~P.~K.; Houshang,~A.; Awad,~A.~A.;
  D{\"u}rrenfeld,~P.; Malm,~B.~G.; Rusu,~A.; {\AA}kerman,~J. \emph{Proc. IEEE}
  \textbf{2016}, \emph{104}, 1919--1945\relax
\mciteBstWouldAddEndPuncttrue
\mciteSetBstMidEndSepPunct{\mcitedefaultmidpunct}
{\mcitedefaultendpunct}{\mcitedefaultseppunct}\relax
\EndOfBibitem
\bibitem[Bychkov and Rashba(1984)Bychkov, and Rashba]{bychkov1984properties}
Bychkov,~Y.~A.; Rashba,~{\'E}.~I. \emph{JETP Lett.} \textbf{1984}, \emph{39},
  78--81\relax
\mciteBstWouldAddEndPuncttrue
\mciteSetBstMidEndSepPunct{\mcitedefaultmidpunct}
{\mcitedefaultendpunct}{\mcitedefaultseppunct}\relax
\EndOfBibitem
\bibitem[Rashba and Sheka(2015)Rashba, and Sheka]{rashba2015symmetry}
Rashba,~E.; Sheka,~V. \emph{New J. Phys.} \textbf{2015}, \emph{17},
  050202\relax
\mciteBstWouldAddEndPuncttrue
\mciteSetBstMidEndSepPunct{\mcitedefaultmidpunct}
{\mcitedefaultendpunct}{\mcitedefaultseppunct}\relax
\EndOfBibitem
\bibitem[Edelstein(1990)]{edelstein1990spin}
Edelstein,~V.~M. \emph{Solid State Commun.} \textbf{1990}, \emph{73},
  233--235\relax
\mciteBstWouldAddEndPuncttrue
\mciteSetBstMidEndSepPunct{\mcitedefaultmidpunct}
{\mcitedefaultendpunct}{\mcitedefaultseppunct}\relax
\EndOfBibitem
\bibitem[Ando and Shiraishi(2017)Ando, and Shiraishi]{ando2017spin}
Ando,~Y.; Shiraishi,~M. \emph{J. Phys. Soc. Jpn.} \textbf{2017}, \emph{86},
  011001\relax
\mciteBstWouldAddEndPuncttrue
\mciteSetBstMidEndSepPunct{\mcitedefaultmidpunct}
{\mcitedefaultendpunct}{\mcitedefaultseppunct}\relax
\EndOfBibitem
\bibitem[Shen \latin{et~al.}(2014)Shen, Vignale, and
  Raimondi]{shen2014microscopic}
Shen,~K.; Vignale,~G.; Raimondi,~R. \emph{Phys. Rev. Lett.} \textbf{2014},
  \emph{112}, 096601\relax
\mciteBstWouldAddEndPuncttrue
\mciteSetBstMidEndSepPunct{\mcitedefaultmidpunct}
{\mcitedefaultendpunct}{\mcitedefaultseppunct}\relax
\EndOfBibitem
\bibitem[S{\'a}nchez \latin{et~al.}(2013)S{\'a}nchez, Vila, Desfonds,
  Gambarelli, Attan{\'e}, De~Teresa, Mag{\'e}n, and Fert]{sanchez2013spin}
S{\'a}nchez,~J.~R.; Vila,~L.; Desfonds,~G.; Gambarelli,~S.; Attan{\'e},~J.;
  De~Teresa,~J.; Mag{\'e}n,~C.; Fert,~A. \emph{Nat. Commun.} \textbf{2013},
  \emph{4}, 2944--2951\relax
\mciteBstWouldAddEndPuncttrue
\mciteSetBstMidEndSepPunct{\mcitedefaultmidpunct}
{\mcitedefaultendpunct}{\mcitedefaultseppunct}\relax
\EndOfBibitem
\bibitem[Manchon \latin{et~al.}(2015)Manchon, Koo, Nitta, Frolov, and
  Duine]{manchon2015new}
Manchon,~A.; Koo,~H.~C.; Nitta,~J.; Frolov,~S.; Duine,~R. \emph{Nat. Mater.}
  \textbf{2015}, \emph{14}, 871--882\relax
\mciteBstWouldAddEndPuncttrue
\mciteSetBstMidEndSepPunct{\mcitedefaultmidpunct}
{\mcitedefaultendpunct}{\mcitedefaultseppunct}\relax
\EndOfBibitem
\bibitem[Rojas-S\'anchez \latin{et~al.}(2016)Rojas-S\'anchez, Oyarz\'un, Fu,
  Marty, Vergnaud, Gambarelli, Vila, Jamet, Ohtsubo, Taleb-Ibrahimi,
  Le~F\`evre, Bertran, Reyren, George, and Fert]{rojas2016spin}
Rojas-S\'anchez,~J.-C.; Oyarz\'un,~S.; Fu,~Y.; Marty,~A.; Vergnaud,~C.;
  Gambarelli,~S.; Vila,~L.; Jamet,~M.; Ohtsubo,~Y.; Taleb-Ibrahimi,~A.;
  Le~F\`evre,~P.; Bertran,~F.; Reyren,~N.; George,~J.-M.; Fert,~A. \emph{Phys.
  Rev. Lett.} \textbf{2016}, \emph{116}, 096602\relax
\mciteBstWouldAddEndPuncttrue
\mciteSetBstMidEndSepPunct{\mcitedefaultmidpunct}
{\mcitedefaultendpunct}{\mcitedefaultseppunct}\relax
\EndOfBibitem
\bibitem[Kurebayashi \latin{et~al.}(2022)Kurebayashi, Garcia, Khan, Sinova, and
  Roche]{kurebayashi2022magnetism}
Kurebayashi,~H.; Garcia,~J.~H.; Khan,~S.; Sinova,~J.; Roche,~S. \emph{Nat. Rev.
  Phys.} \textbf{2022}, \emph{4}, 150--166\relax
\mciteBstWouldAddEndPuncttrue
\mciteSetBstMidEndSepPunct{\mcitedefaultmidpunct}
{\mcitedefaultendpunct}{\mcitedefaultseppunct}\relax
\EndOfBibitem
\bibitem[Vaz \latin{et~al.}(2019)Vaz, No{\"e}l, Johansson, G{\"o}bel, Bruno,
  Singh, Mckeown-Walker, Trier, Vicente-Arche, Sander, Valencia, Bruneel,
  Vivek, Gabay, Bergeal, Baumberger, Okuno, Barthélémy, Fert, Vila, Mertig,
  Attané, and Bibes]{vaz2019mapping}
Vaz,~D.~C. \latin{et~al.}  \emph{Nat. Mater.} \textbf{2019}, \emph{18},
  1187--1193\relax
\mciteBstWouldAddEndPuncttrue
\mciteSetBstMidEndSepPunct{\mcitedefaultmidpunct}
{\mcitedefaultendpunct}{\mcitedefaultseppunct}\relax
\EndOfBibitem
\bibitem[Rezende(2020)]{rezende2020fundamentals}
Rezende,~S.~M. \emph{{Fundamentals of Magnonics}}; Springer, 2020; Vol.
  969\relax
\mciteBstWouldAddEndPuncttrue
\mciteSetBstMidEndSepPunct{\mcitedefaultmidpunct}
{\mcitedefaultendpunct}{\mcitedefaultseppunct}\relax
\EndOfBibitem
\bibitem[Liu \latin{et~al.}(2013)Liu, Shan, Yao, Yao, and Xiao]{liu2013three}
Liu,~G.-B.; Shan,~W.-Y.; Yao,~Y.; Yao,~W.; Xiao,~D. \emph{Phys. Rev. B}
  \textbf{2013}, \emph{88}, 085433\relax
\mciteBstWouldAddEndPuncttrue
\mciteSetBstMidEndSepPunct{\mcitedefaultmidpunct}
{\mcitedefaultendpunct}{\mcitedefaultseppunct}\relax
\EndOfBibitem
\bibitem[Sangiao \latin{et~al.}(2015)Sangiao, De~Teresa, Morellon, Lucas,
  Mart{\'\i}nez-Velarte, and Viret]{sangiao2015control}
Sangiao,~S.; De~Teresa,~J.; Morellon,~L.; Lucas,~I.;
  Mart{\'\i}nez-Velarte,~M.~C.; Viret,~M. \emph{Appl. Phys. Lett.}
  \textbf{2015}, \emph{106}, 172403\relax
\mciteBstWouldAddEndPuncttrue
\mciteSetBstMidEndSepPunct{\mcitedefaultmidpunct}
{\mcitedefaultendpunct}{\mcitedefaultseppunct}\relax
\EndOfBibitem
\bibitem[Zhang \latin{et~al.}(2015)Zhang, Jungfleisch, Jiang, Pearson, and
  Hoffmann]{zhang2015spin}
Zhang,~W.; Jungfleisch,~M.~B.; Jiang,~W.; Pearson,~J.~E.; Hoffmann,~A. \emph{J.
  Appl. Phys.} \textbf{2015}, \emph{117}, 17C727\relax
\mciteBstWouldAddEndPuncttrue
\mciteSetBstMidEndSepPunct{\mcitedefaultmidpunct}
{\mcitedefaultendpunct}{\mcitedefaultseppunct}\relax
\EndOfBibitem
\bibitem[Isasa \latin{et~al.}(2016)Isasa, Mart\'{\i}nez-Velarte, Villamor,
  Mag\'en, Morell\'on, De~Teresa, Ibarra, Vignale, Chulkov, Krasovskii, Hueso,
  and Casanova]{isasa2016origin}
Isasa,~M.; Mart\'{\i}nez-Velarte,~M.~C.; Villamor,~E.; Mag\'en,~C.;
  Morell\'on,~L.; De~Teresa,~J.~M.; Ibarra,~M.~R.; Vignale,~G.; Chulkov,~E.~V.;
  Krasovskii,~E.~E.; Hueso,~L.~E.; Casanova,~F. \emph{Phys. Rev. B}
  \textbf{2016}, \emph{93}, 014420\relax
\mciteBstWouldAddEndPuncttrue
\mciteSetBstMidEndSepPunct{\mcitedefaultmidpunct}
{\mcitedefaultendpunct}{\mcitedefaultseppunct}\relax
\EndOfBibitem
\bibitem[Nomura \latin{et~al.}(2015)Nomura, Tashiro, Nakayama, and
  Ando]{nomura2015temperature}
Nomura,~A.; Tashiro,~T.; Nakayama,~H.; Ando,~K. \emph{Appl. Phys. Lett.}
  \textbf{2015}, \emph{106}, 212403\relax
\mciteBstWouldAddEndPuncttrue
\mciteSetBstMidEndSepPunct{\mcitedefaultmidpunct}
{\mcitedefaultendpunct}{\mcitedefaultseppunct}\relax
\EndOfBibitem
\bibitem[Karube \latin{et~al.}(2016)Karube, Kondou, and
  Otani]{karube2016experimental}
Karube,~S.; Kondou,~K.; Otani,~Y. \emph{Appl. Phys. Express} \textbf{2016},
  \emph{9}, 033001\relax
\mciteBstWouldAddEndPuncttrue
\mciteSetBstMidEndSepPunct{\mcitedefaultmidpunct}
{\mcitedefaultendpunct}{\mcitedefaultseppunct}\relax
\EndOfBibitem
\bibitem[Zhou \latin{et~al.}(2018)Zhou, Liu, Wang, Ma, Jia, Wu, Zhou, Zhang,
  Liu, Wu, and Qi]{zhou2018broadband}
Zhou,~C.; Liu,~Y.~P.; Wang,~Z.; Ma,~S.~J.; Jia,~M.~W.; Wu,~R.~Q.; Zhou,~L.;
  Zhang,~W.; Liu,~M.~K.; Wu,~Y.~Z.; Qi,~J. \emph{Phys. Rev. Lett.}
  \textbf{2018}, \emph{121}, 086801\relax
\mciteBstWouldAddEndPuncttrue
\mciteSetBstMidEndSepPunct{\mcitedefaultmidpunct}
{\mcitedefaultendpunct}{\mcitedefaultseppunct}\relax
\EndOfBibitem
\bibitem[Ohshima \latin{et~al.}(2014)Ohshima, Sakai, Ando, Shinjo, Kawahara,
  Ago, and Shiraishi]{ohshima2014observation}
Ohshima,~R.; Sakai,~A.; Ando,~Y.; Shinjo,~T.; Kawahara,~K.; Ago,~H.;
  Shiraishi,~M. \emph{Appl. Phys. Lett.} \textbf{2014}, \emph{105},
  162410\relax
\mciteBstWouldAddEndPuncttrue
\mciteSetBstMidEndSepPunct{\mcitedefaultmidpunct}
{\mcitedefaultendpunct}{\mcitedefaultseppunct}\relax
\EndOfBibitem
\bibitem[Mendes \latin{et~al.}(2015)Mendes, Santos, Meireles, Lacerda,
  Vilela-Le{\~a}o, Machado, Rodr{\'\i}guez-Su{\'a}rez, Azevedo, and
  Rezende]{mendes2015spin}
Mendes,~J.; Santos,~O.~A.; Meireles,~L.; Lacerda,~R.; Vilela-Le{\~a}o,~L.;
  Machado,~F.; Rodr{\'\i}guez-Su{\'a}rez,~R.; Azevedo,~A.; Rezende,~S.
  \emph{Phys. Rev. Lett.} \textbf{2015}, \emph{115}, 226601\relax
\mciteBstWouldAddEndPuncttrue
\mciteSetBstMidEndSepPunct{\mcitedefaultmidpunct}
{\mcitedefaultendpunct}{\mcitedefaultseppunct}\relax
\EndOfBibitem
\bibitem[Gusakova \latin{et~al.}(2017)Gusakova, Wang, Shiau, Krivosheeva,
  Shaposhnikov, Borisenko, Gusakov, and Tay]{gusakova2017electronic}
Gusakova,~J.; Wang,~X.; Shiau,~L.~L.; Krivosheeva,~A.; Shaposhnikov,~V.;
  Borisenko,~V.; Gusakov,~V.; Tay,~B.~K. \emph{Phys. Status Solidi A}
  \textbf{2017}, \emph{214}, 1700218\relax
\mciteBstWouldAddEndPuncttrue
\mciteSetBstMidEndSepPunct{\mcitedefaultmidpunct}
{\mcitedefaultendpunct}{\mcitedefaultseppunct}\relax
\EndOfBibitem
\bibitem[Mendes \latin{et~al.}(2018)Mendes, Aparecido-Ferreira, Holanda,
  Azevedo, and Rezende]{mendes2018efficient}
Mendes,~J.; Aparecido-Ferreira,~A.; Holanda,~J.; Azevedo,~A.; Rezende,~S.
  \emph{Appl. Phys. Lett.} \textbf{2018}, \emph{112}, 242407\relax
\mciteBstWouldAddEndPuncttrue
\mciteSetBstMidEndSepPunct{\mcitedefaultmidpunct}
{\mcitedefaultendpunct}{\mcitedefaultseppunct}\relax
\EndOfBibitem
\bibitem[Bhowal and Satpathy(2020)Bhowal, and Satpathy]{bhowal2020intrinsic}
Bhowal,~S.; Satpathy,~S. \emph{Phys. Rev. B} \textbf{2020}, \emph{102},
  035409\relax
\mciteBstWouldAddEndPuncttrue
\mciteSetBstMidEndSepPunct{\mcitedefaultmidpunct}
{\mcitedefaultendpunct}{\mcitedefaultseppunct}\relax
\EndOfBibitem
\bibitem[Shao \latin{et~al.}(2016)Shao, Yu, Lan, Shi, Li, Zheng, Zhu, Li,
  Amiri, and Wang]{shao2016strong}
Shao,~Q.; Yu,~G.; Lan,~Y.-W.; Shi,~Y.; Li,~M.-Y.; Zheng,~C.; Zhu,~X.;
  Li,~L.-J.; Amiri,~P.~K.; Wang,~K.~L. \emph{Nano Lett.} \textbf{2016},
  \emph{16}, 7514--7520\relax
\mciteBstWouldAddEndPuncttrue
\mciteSetBstMidEndSepPunct{\mcitedefaultmidpunct}
{\mcitedefaultendpunct}{\mcitedefaultseppunct}\relax
\EndOfBibitem
\bibitem[Zhao \latin{et~al.}(2020)Zhao, Karpiak, Khokhriakov, Johansson, Hoque,
  Xu, Jiang, Mertig, and Dash]{zhao2020unconventional}
Zhao,~B.; Karpiak,~B.; Khokhriakov,~D.; Johansson,~A.; Hoque,~A.~M.; Xu,~X.;
  Jiang,~Y.; Mertig,~I.; Dash,~S.~P. \emph{Adv. Mater.} \textbf{2020},
  \emph{32}, 2000818\relax
\mciteBstWouldAddEndPuncttrue
\mciteSetBstMidEndSepPunct{\mcitedefaultmidpunct}
{\mcitedefaultendpunct}{\mcitedefaultseppunct}\relax
\EndOfBibitem
\bibitem[MacNeill \latin{et~al.}(2017)MacNeill, Stiehl, Guimaraes, Buhrman,
  Park, and Ralph]{macneill2017control}
MacNeill,~D.; Stiehl,~G.; Guimaraes,~M.; Buhrman,~R.; Park,~J.; Ralph,~D.
  \emph{Nat. Phys.} \textbf{2017}, \emph{13}, 300--305\relax
\mciteBstWouldAddEndPuncttrue
\mciteSetBstMidEndSepPunct{\mcitedefaultmidpunct}
{\mcitedefaultendpunct}{\mcitedefaultseppunct}\relax
\EndOfBibitem
\bibitem[Lee \latin{et~al.}(2020)Lee, Lee, Kikkawa, Le, Kang, Kim, Nguyen, Kim,
  Park, and Saitoh]{lee2020enhanced}
Lee,~S.-K.; Lee,~W.-Y.; Kikkawa,~T.; Le,~C.~T.; Kang,~M.-S.; Kim,~G.-S.;
  Nguyen,~A.~D.; Kim,~Y.~S.; Park,~N.-W.; Saitoh,~E. \emph{Adv. Funct. Mater.}
  \textbf{2020}, \emph{30}, 2003192\relax
\mciteBstWouldAddEndPuncttrue
\mciteSetBstMidEndSepPunct{\mcitedefaultmidpunct}
{\mcitedefaultendpunct}{\mcitedefaultseppunct}\relax
\EndOfBibitem
\bibitem[Lee \latin{et~al.}(2020)Lee, Park, Kim, Kang, Choi, Choi, Jang,
  Saitoh, and Lee]{lee2020enhanced2}
Lee,~W.-Y.; Park,~N.-W.; Kim,~G.-S.; Kang,~M.-S.; Choi,~J.~W.; Choi,~K.-Y.;
  Jang,~H.~W.; Saitoh,~E.; Lee,~S.-K. \emph{Nano Lett.} \textbf{2020},
  \emph{21}, 189--196\relax
\mciteBstWouldAddEndPuncttrue
\mciteSetBstMidEndSepPunct{\mcitedefaultmidpunct}
{\mcitedefaultendpunct}{\mcitedefaultseppunct}\relax
\EndOfBibitem
\bibitem[Dastgeer \latin{et~al.}(2019)Dastgeer, Shehzad, and
  Eom]{dastgeer2019distinct}
Dastgeer,~G.; Shehzad,~M.~A.; Eom,~J. \emph{ACS Appl. Mater. Interfaces}
  \textbf{2019}, \emph{11}, 48533--48539\relax
\mciteBstWouldAddEndPuncttrue
\mciteSetBstMidEndSepPunct{\mcitedefaultmidpunct}
{\mcitedefaultendpunct}{\mcitedefaultseppunct}\relax
\EndOfBibitem
\bibitem[Hoque \latin{et~al.}(2020)Hoque, Khokhriakov, Karpiak, and
  Dash]{PhysRevResearch.2.033204}
Hoque,~A.~M.; Khokhriakov,~D.; Karpiak,~B.; Dash,~S.~P. \emph{Phys. Rev.
  Research} \textbf{2020}, \emph{2}, 033204\relax
\mciteBstWouldAddEndPuncttrue
\mciteSetBstMidEndSepPunct{\mcitedefaultmidpunct}
{\mcitedefaultendpunct}{\mcitedefaultseppunct}\relax
\EndOfBibitem
\bibitem[Zhao \latin{et~al.}(2020)Zhao, Khokhriakov, Zhang, Fu, Karpiak, Hoque,
  Xu, Jiang, Yan, and Dash]{PhysRevResearch.2.013286}
Zhao,~B.; Khokhriakov,~D.; Zhang,~Y.; Fu,~H.; Karpiak,~B.; Hoque,~A.~M.;
  Xu,~X.; Jiang,~Y.; Yan,~B.; Dash,~S.~P. \emph{Phys. Rev. Research}
  \textbf{2020}, \emph{2}, 013286\relax
\mciteBstWouldAddEndPuncttrue
\mciteSetBstMidEndSepPunct{\mcitedefaultmidpunct}
{\mcitedefaultendpunct}{\mcitedefaultseppunct}\relax
\EndOfBibitem
\bibitem[Kumar \latin{et~al.}(2020)Kumar, Chaurasiya, Chowdhury, Mondal,
  Bansal, Barvat, Khanna, Pal, Chaudhary, Barman, and Muduli]{kumar2020direct}
Kumar,~A.; Chaurasiya,~A.~K.; Chowdhury,~N.; Mondal,~A.~K.; Bansal,~R.;
  Barvat,~A.; Khanna,~S.~P.; Pal,~P.; Chaudhary,~S.; Barman,~A.; Muduli,~P.~K.
  \emph{Appl. Phys. Lett.} \textbf{2020}, \emph{116}, 232405\relax
\mciteBstWouldAddEndPuncttrue
\mciteSetBstMidEndSepPunct{\mcitedefaultmidpunct}
{\mcitedefaultendpunct}{\mcitedefaultseppunct}\relax
\EndOfBibitem
\bibitem[Bansal \latin{et~al.}(2019)Bansal, Kumar, Chowdhury, Sisodia, Barvat,
  Dogra, Pal, and Muduli]{bansal2019extrinsic}
Bansal,~R.; Kumar,~A.; Chowdhury,~N.; Sisodia,~N.; Barvat,~A.; Dogra,~A.;
  Pal,~P.; Muduli,~P. \emph{J. Magn. Magn. Mater.} \textbf{2019}, \emph{476},
  337--341\relax
\mciteBstWouldAddEndPuncttrue
\mciteSetBstMidEndSepPunct{\mcitedefaultmidpunct}
{\mcitedefaultendpunct}{\mcitedefaultseppunct}\relax
\EndOfBibitem
\bibitem[Shi \latin{et~al.}(2018)Shi, Yang, Jiang, Zhang, Huan, Xie, Hong, Shi,
  and Zhang]{shi2018assisted}
Shi,~Y.; Yang,~P.; Jiang,~S.; Zhang,~Z.; Huan,~Y.; Xie,~C.; Hong,~M.; Shi,~J.;
  Zhang,~Y. \emph{Nanotechnology} \textbf{2018}, \emph{30}, 034002\relax
\mciteBstWouldAddEndPuncttrue
\mciteSetBstMidEndSepPunct{\mcitedefaultmidpunct}
{\mcitedefaultendpunct}{\mcitedefaultseppunct}\relax
\EndOfBibitem
\bibitem[Berkdemir \latin{et~al.}(2013)Berkdemir, Guti{\'e}rrez,
  Botello-M{\'e}ndez, Perea-L{\'o}pez, El{\'\i}as, Chia, Wang, Crespi,
  L{\'o}pez-Ur{\'\i}as, Charlier, Terrones, and
  Terrones]{berkdemir2013identification}
Berkdemir,~A.; Guti{\'e}rrez,~H.~R.; Botello-M{\'e}ndez,~A.~R.;
  Perea-L{\'o}pez,~N.; El{\'\i}as,~A.~L.; Chia,~C.-I.; Wang,~B.; Crespi,~V.~H.;
  L{\'o}pez-Ur{\'\i}as,~F.; Charlier,~J.-C.; Terrones,~H.; Terrones,~M.
  \emph{Sci. Rep.} \textbf{2013}, \emph{3}, 1755\relax
\mciteBstWouldAddEndPuncttrue
\mciteSetBstMidEndSepPunct{\mcitedefaultmidpunct}
{\mcitedefaultendpunct}{\mcitedefaultseppunct}\relax
\EndOfBibitem
\bibitem[Tonndorf \latin{et~al.}(2013)Tonndorf, Schmidt, B\"{o}ttger, Zhang,
  B\"{o}rner, Liebig, Albrecht, Kloc, Gordan, Zahn, de~Vasconcellos, and
  Bratschitsch]{tonndorf2013photoluminescence}
Tonndorf,~P.; Schmidt,~R.; B\"{o}ttger,~P.; Zhang,~X.; B\"{o}rner,~J.;
  Liebig,~A.; Albrecht,~M.; Kloc,~C.; Gordan,~O.; Zahn,~D. R.~T.;
  de~Vasconcellos,~S.~M.; Bratschitsch,~R. \emph{Opt. Express} \textbf{2013},
  \emph{21}, 4908--4916\relax
\mciteBstWouldAddEndPuncttrue
\mciteSetBstMidEndSepPunct{\mcitedefaultmidpunct}
{\mcitedefaultendpunct}{\mcitedefaultseppunct}\relax
\EndOfBibitem
\bibitem[Mahjouri-Samani \latin{et~al.}(2016)Mahjouri-Samani, Liang, Oyedele,
  Kim, Tian, Cross, Wang, Lin, Boulesbaa, Rouleau, Puretzky, Xiao, Yoon, Eres,
  Duscher, Sumpter, and Geohegan]{mahjouri2016tailoring}
Mahjouri-Samani,~M. \latin{et~al.}  \emph{Nano Lett.} \textbf{2016}, \emph{16},
  5213--5220\relax
\mciteBstWouldAddEndPuncttrue
\mciteSetBstMidEndSepPunct{\mcitedefaultmidpunct}
{\mcitedefaultendpunct}{\mcitedefaultseppunct}\relax
\EndOfBibitem
\bibitem[Chaurasiya \latin{et~al.}(2019)Chaurasiya, Kumar, Gupta, Chaudhary,
  Muduli, and Barman]{chaurasiya2019direct}
Chaurasiya,~A.~K.; Kumar,~A.; Gupta,~R.; Chaudhary,~S.; Muduli,~P.~K.;
  Barman,~A. \emph{Phys. Rev. B} \textbf{2019}, \emph{99}, 035402\relax
\mciteBstWouldAddEndPuncttrue
\mciteSetBstMidEndSepPunct{\mcitedefaultmidpunct}
{\mcitedefaultendpunct}{\mcitedefaultseppunct}\relax
\EndOfBibitem
\bibitem[Kumar \latin{et~al.}(2018)Kumar, Bansal, Chaudhary, and
  Muduli]{kumar2018large}
Kumar,~A.; Bansal,~R.; Chaudhary,~S.; Muduli,~P.~K. \emph{Phys. Rev. B}
  \textbf{2018}, \emph{98}, 104403\relax
\mciteBstWouldAddEndPuncttrue
\mciteSetBstMidEndSepPunct{\mcitedefaultmidpunct}
{\mcitedefaultendpunct}{\mcitedefaultseppunct}\relax
\EndOfBibitem
\bibitem[Lustikova \latin{et~al.}(2015)Lustikova, Shiomi, and
  Saitoh]{lustikova2015vector}
Lustikova,~J.; Shiomi,~Y.; Saitoh,~E. \emph{Phys. Rev. B} \textbf{2015},
  \emph{92}, 224436\relax
\mciteBstWouldAddEndPuncttrue
\mciteSetBstMidEndSepPunct{\mcitedefaultmidpunct}
{\mcitedefaultendpunct}{\mcitedefaultseppunct}\relax
\EndOfBibitem
\bibitem[Cerqueira \latin{et~al.}(2018)Cerqueira, Qin, Dang, Djeffal,
  Le~Breton, Hehn, Rojas-Sanchez, Devaux, Suire, Migot, \latin{et~al.}
  others]{cerqueira2018evidence}
Cerqueira,~C.; Qin,~J.~Y.; Dang,~H.; Djeffal,~A.; Le~Breton,~J.-C.; Hehn,~M.;
  Rojas-Sanchez,~J.-C.; Devaux,~X.; Suire,~S.; Migot,~S., \latin{et~al.}
  \emph{Nano Lett.} \textbf{2018}, \emph{19}, 90--99\relax
\mciteBstWouldAddEndPuncttrue
\mciteSetBstMidEndSepPunct{\mcitedefaultmidpunct}
{\mcitedefaultendpunct}{\mcitedefaultseppunct}\relax
\EndOfBibitem
\bibitem[Gladii \latin{et~al.}(2019)Gladii, Frangou, Hallal, Seeger, No\"el,
  Forestier, Auffret, Rubio-Roy, Warin, Vila, Wimmer, Ebert, Gambarelli,
  Chshiev, and Baltz]{gladii2019self}
Gladii,~O.; Frangou,~L.; Hallal,~A.; Seeger,~R.~L.; No\"el,~P.; Forestier,~G.;
  Auffret,~S.; Rubio-Roy,~M.; Warin,~P.; Vila,~L.; Wimmer,~S.; Ebert,~H.;
  Gambarelli,~S.; Chshiev,~M.; Baltz,~V. \emph{Phys. Rev. B} \textbf{2019},
  \emph{100}, 174409\relax
\mciteBstWouldAddEndPuncttrue
\mciteSetBstMidEndSepPunct{\mcitedefaultmidpunct}
{\mcitedefaultendpunct}{\mcitedefaultseppunct}\relax
\EndOfBibitem
\bibitem[Villamor \latin{et~al.}(2013)Villamor, Isasa, Hueso, and
  Casanova]{villamor2013contribution}
Villamor,~E.; Isasa,~M.; Hueso,~L.~E.; Casanova,~F. \emph{Phys. Rev. B}
  \textbf{2013}, \emph{87}, 094417\relax
\mciteBstWouldAddEndPuncttrue
\mciteSetBstMidEndSepPunct{\mcitedefaultmidpunct}
{\mcitedefaultendpunct}{\mcitedefaultseppunct}\relax
\EndOfBibitem
\bibitem[Muduli \latin{et~al.}(2018)Muduli, Kimata, Omori, Wakamura, Dash, and
  Otani]{PhysRevB.98.024416}
Muduli,~P.~K.; Kimata,~M.; Omori,~Y.; Wakamura,~T.; Dash,~S.~P.; Otani,~Y.
  \emph{Phys. Rev. B} \textbf{2018}, \emph{98}, 024416\relax
\mciteBstWouldAddEndPuncttrue
\mciteSetBstMidEndSepPunct{\mcitedefaultmidpunct}
{\mcitedefaultendpunct}{\mcitedefaultseppunct}\relax
\EndOfBibitem
\bibitem[Tserkovnyak \latin{et~al.}(2002)Tserkovnyak, Brataas, and
  Bauer]{tserkovnyak2002enhanced}
Tserkovnyak,~Y.; Brataas,~A.; Bauer,~G.~E. \emph{Phys. Rev. Lett.}
  \textbf{2002}, \emph{88}, 117601\relax
\mciteBstWouldAddEndPuncttrue
\mciteSetBstMidEndSepPunct{\mcitedefaultmidpunct}
{\mcitedefaultendpunct}{\mcitedefaultseppunct}\relax
\EndOfBibitem
\bibitem[Kittel(1948)]{kittel1948theory}
Kittel,~C. \emph{Phys. Rev.} \textbf{1948}, \emph{73}, 155--161\relax
\mciteBstWouldAddEndPuncttrue
\mciteSetBstMidEndSepPunct{\mcitedefaultmidpunct}
{\mcitedefaultendpunct}{\mcitedefaultseppunct}\relax
\EndOfBibitem
\bibitem[Tyer \latin{et~al.}(2003)Tyer, Van~der Laan, Temmerman, Szotek, and
  Ebert]{tyer2003systematic}
Tyer,~R.; Van~der Laan,~G.; Temmerman,~W.; Szotek,~Z.; Ebert,~H. \emph{Phys.
  Rev. B} \textbf{2003}, \emph{67}, 104409\relax
\mciteBstWouldAddEndPuncttrue
\mciteSetBstMidEndSepPunct{\mcitedefaultmidpunct}
{\mcitedefaultendpunct}{\mcitedefaultseppunct}\relax
\EndOfBibitem
\bibitem[Tsai \latin{et~al.}(2018)Tsai, Karube, Kondou, Yamaguchi, Ishii, and
  Otani]{tsai2018clear}
Tsai,~H.; Karube,~S.; Kondou,~K.; Yamaguchi,~N.; Ishii,~F.; Otani,~Y.
  \emph{Sci. Rep.} \textbf{2018}, \emph{8}, 5564--5572\relax
\mciteBstWouldAddEndPuncttrue
\mciteSetBstMidEndSepPunct{\mcitedefaultmidpunct}
{\mcitedefaultendpunct}{\mcitedefaultseppunct}\relax
\EndOfBibitem
\bibitem[Liang \latin{et~al.}(2017)Liang, Yang, Renucci, Tao, Laczkowski,
  Mc-Murtry, Wang, Marie, George, Petit-Watelot, Djeffal, Mangin, Jaffrès, and
  Lu]{liang2017electrical}
Liang,~S.; Yang,~H.; Renucci,~P.; Tao,~B.; Laczkowski,~P.; Mc-Murtry,~S.;
  Wang,~G.; Marie,~X.; George,~J.-M.; Petit-Watelot,~S.; Djeffal,~A.;
  Mangin,~S.; Jaffrès,~H.; Lu,~Y. \emph{Nat. Commun.} \textbf{2017}, \emph{8},
  14947\relax
\mciteBstWouldAddEndPuncttrue
\mciteSetBstMidEndSepPunct{\mcitedefaultmidpunct}
{\mcitedefaultendpunct}{\mcitedefaultseppunct}\relax
\EndOfBibitem
\bibitem[Chang \latin{et~al.}(2014)Chang, Zhang, Zhu, Han, Pu, Chang, Hsu,
  Huang, Hsu, Chiu, Takenobu, Li, Wu, Chang, Wee, and Li]{chang2014monolayer}
Chang,~Y.-H. \latin{et~al.}  \emph{ACS Nano} \textbf{2014}, \emph{8},
  8582--8590\relax
\mciteBstWouldAddEndPuncttrue
\mciteSetBstMidEndSepPunct{\mcitedefaultmidpunct}
{\mcitedefaultendpunct}{\mcitedefaultseppunct}\relax
\EndOfBibitem
\bibitem[Zhang \latin{et~al.}(2013)Zhang, Zhang, Ji, Ju, Yuan, Shi, Gao, Ma,
  Liu, Chen, Song, Hwang, Cui, and Liu]{zhang2013controlled}
Zhang,~Y.; Zhang,~Y.; Ji,~Q.; Ju,~J.; Yuan,~H.; Shi,~J.; Gao,~T.; Ma,~D.;
  Liu,~M.; Chen,~Y.; Song,~X.; Hwang,~H.~Y.; Cui,~Y.; Liu,~Z. \emph{ACS Nano}
  \textbf{2013}, \emph{7}, 8963--8971\relax
\mciteBstWouldAddEndPuncttrue
\mciteSetBstMidEndSepPunct{\mcitedefaultmidpunct}
{\mcitedefaultendpunct}{\mcitedefaultseppunct}\relax
\EndOfBibitem
\bibitem[Zhang \latin{et~al.}(2015)Zhang, Han, Jiang, Yang, and
  Parkin]{zhang2015role}
Zhang,~W.; Han,~W.; Jiang,~X.; Yang,~S.-H.; Parkin,~S.~S. \emph{Nat. Phys.}
  \textbf{2015}, \emph{11}, 496--502\relax
\mciteBstWouldAddEndPuncttrue
\mciteSetBstMidEndSepPunct{\mcitedefaultmidpunct}
{\mcitedefaultendpunct}{\mcitedefaultseppunct}\relax
\EndOfBibitem
\bibitem[Hoffmann(2013)]{hoffmann2013spin}
Hoffmann,~A. \emph{IEEE Trans. Magn.} \textbf{2013}, \emph{49},
  5172--5193\relax
\mciteBstWouldAddEndPuncttrue
\mciteSetBstMidEndSepPunct{\mcitedefaultmidpunct}
{\mcitedefaultendpunct}{\mcitedefaultseppunct}\relax
\EndOfBibitem
\bibitem[Tiwari \latin{et~al.}(2017)Tiwari, Behera, Kumar, D{\"u}rrenfeld,
  Chaudhary, Pandya, {\AA}kerman, and Muduli]{tiwari2017antidamping}
Tiwari,~D.; Behera,~N.; Kumar,~A.; D{\"u}rrenfeld,~P.; Chaudhary,~S.;
  Pandya,~D.; {\AA}kerman,~J.; Muduli,~P.~K. \emph{Appl. Phys. Lett.}
  \textbf{2017}, \emph{111}, 232407\relax
\mciteBstWouldAddEndPuncttrue
\mciteSetBstMidEndSepPunct{\mcitedefaultmidpunct}
{\mcitedefaultendpunct}{\mcitedefaultseppunct}\relax
\EndOfBibitem
\bibitem[Bansal \latin{et~al.}(2018)Bansal, Nirala, Kumar, Chaudhary, and
  Muduli]{bansal2018large}
Bansal,~R.; Nirala,~G.; Kumar,~A.; Chaudhary,~S.; Muduli,~P.~K. \emph{Spin}
  \textbf{2018}, \emph{8}, 1850018\relax
\mciteBstWouldAddEndPuncttrue
\mciteSetBstMidEndSepPunct{\mcitedefaultmidpunct}
{\mcitedefaultendpunct}{\mcitedefaultseppunct}\relax
\EndOfBibitem
\bibitem[Yakata \latin{et~al.}(2006)Yakata, Ando, Miyazaki, and
  Mizukami]{yakata2006temperature}
Yakata,~S.; Ando,~Y.; Miyazaki,~T.; Mizukami,~S. \emph{Jpn. J. Appl. Phys.}
  \textbf{2006}, \emph{45}, 3892--3895\relax
\mciteBstWouldAddEndPuncttrue
\mciteSetBstMidEndSepPunct{\mcitedefaultmidpunct}
{\mcitedefaultendpunct}{\mcitedefaultseppunct}\relax
\EndOfBibitem
\bibitem[Yu \latin{et~al.}(2018)Yu, Miao, Sun, Liu, Du, Omelchenko, Heinrich,
  Wu, and Ding]{yu2018determination}
Yu,~R.; Miao,~B.~F.; Sun,~L.; Liu,~Q.; Du,~J.; Omelchenko,~P.; Heinrich,~B.;
  Wu,~M.; Ding,~H.~F. \emph{Phys. Rev. Materials} \textbf{2018}, \emph{2},
  074406\relax
\mciteBstWouldAddEndPuncttrue
\mciteSetBstMidEndSepPunct{\mcitedefaultmidpunct}
{\mcitedefaultendpunct}{\mcitedefaultseppunct}\relax
\EndOfBibitem
\bibitem[Tsukahara \latin{et~al.}(2014)Tsukahara, Ando, Kitamura, Emoto,
  Shikoh, Delmo, Shinjo, and Shiraishi]{tsukahara2014self}
Tsukahara,~A.; Ando,~Y.; Kitamura,~Y.; Emoto,~H.; Shikoh,~E.; Delmo,~M.~P.;
  Shinjo,~T.; Shiraishi,~M. \emph{Phys. Rev. B} \textbf{2014}, \emph{89},
  235317\relax
\mciteBstWouldAddEndPuncttrue
\mciteSetBstMidEndSepPunct{\mcitedefaultmidpunct}
{\mcitedefaultendpunct}{\mcitedefaultseppunct}\relax
\EndOfBibitem
\end{mcitethebibliography}

\end{document}